\documentclass[review,12pt,3p,times]{elsarticle}

\usepackage{amssymb}
\usepackage{lineno}
\usepackage{setspace}
\usepackage[utf8]{inputenc}
\usepackage{amsmath}
\usepackage{overpic}

\usepackage{booktabs}
\usepackage{bm}
\usepackage{comment}
\usepackage{float}
\usepackage{gensymb}
\usepackage{graphicx}
\usepackage{multirow}
\usepackage{nomencl}
\usepackage{natbib}
\usepackage{tikz}
\usepackage{ulem}
\setcitestyle{square,comma,numbers,sort&compress}
\biboptions{numbers,sort&compress}

\usepackage[version=4]{mhchem}
\usepackage{siunitx}
\usepackage{subfigure}
\usepackage{subcaption}
\usepackage{textcomp}
\usepackage{longtable,tabularx}
\usepackage{ulem}
\usepackage{verbatim}
\usepackage{xcolor}
\usepackage{soul}
\usepackage{xpatch}
\usepackage{tcolorbox}

\makenomenclature
\usepackage{etoolbox}
\renewcommand\nomgroup[1]{%
  \item[\bfseries
  \ifstrequal{#1}{A}{}{%
  \ifstrequal{#1}{B}{Greek symbols}{%
  \ifstrequal{#1}{C}{Abbreviations}{}}}%
]}
\begin{document}

\begin{frontmatter}



\title{A novel data-driven method for augmenting turbulence modelling for unsteady cavitating flows }

\author[inst1,inst2]{Dhruv Apte*}
\author[inst3]{Nassim Razaaly}
\author[inst4]{Yuan Fang}
\author[inst5]{Mingming Ge}
\author[inst4]{Richard Sandberg}
\author[inst1,inst6]{Olivier Coutier-Delgosha}

\affiliation[inst1]{organization={Kevin T. Crofton Department of Aerospace and Ocean Engineering},%Department and Organization
            addressline={Virginia Tech}, 
            city={Blacksburg},
            postcode={24060}, 
            state={VA},
            country={USA}}
\affiliation[inst3]{organization={Institut PPrime, ISAE-ENSMA},%Department and Organization
            city={Chasseneuil-du-Poitou},
            %postcode={26507}, 
            country={France}}
\affiliation[inst4]{organization={Department of Mechanical Engineering, University of Melbourne},%Department and Organization
            addressline={Parkville}, 
            city={Victoria},
            postcode={3010},
            country={Australia}}  
\affiliation[inst5]{organization={National Observation and Research Station of Coastal Ecological Environments in Macao, Macao Environmental Research Institute},%Department and Organization
            addressline={Faculty of Innovation Engineering, Macau University of Science and Technology}, 
            city={Macau SAR},
            postcode={999078},
            country={China}}  
\affiliation[inst6]{organization={Univ. Lille, CNRS, ONERA, Arts et Metiers ParisTech},%Department and Organization
            addressline={Centrale Lille, FRE 2017 - LMFL - Laboratoire de Mecanique des fluides de Lille}, 
            city={Kampe de Feriet},
            postcode={F-59000}, 
            state={Lille},
            country={France}}  
\affiliation[inst2]{organization={Oden Institute for Computational Engineering and Sciences},%Department and Organization
            addressline={University of Texas at Austin}, 
            city={Austin},
            postcode={78705}, 
            state={Texas},
            country={USA}}              

\begin{abstract}
%Cavitation is a highly turbulent, multi-phase flow phenomenon that manifests as a result of a sudden drop in the liquid pressure in the form of vapor bubbles. Modelling cavitating flows has been extremely challenging owing to flow unsteadiness, phase transition and the cavitation-turbulence interaction. Standard methods to model turbulence like Reynolds-Averaged Navier-Stokes equations (RANS) and hybrid RANS-Large Eddy Simulations (LES) models are unable to reproduce the local turbulence dynamics observed in experiments. To facilitate the development of accurate RANS models, a data-driven approach is investigated by integrating gene-expression programming (GEP) to obtain a corrective function in the Reynolds stress tensor. In addition, numerical optimization techniques are employed to ensure the solution is independent of the uncertainties associated with GEP. The method is utilized to simulate unsteady cavitating flow in a converging-diverging nozzle (\textit{venturi}). The approach is able to demonstrate the flow dynamics not captured by numerical simulations and thus, projects as a promising approach to augment numerical simulations.
Cavitation is a highly turbulent, multi-phase flow phenomenon that manifests in the form of vapor cavities as a result of a sudden drop in the liquid pressure. The phenomenon has been observed as widely detrimental in hydraulic and marine applications like propellers and pumps, while also accelerating the sub-processes involved for bio-diesel production.  Modelling cavitating flows has been extremely challenging owing to the resulting flow unsteadiness, phase transition and the cavitation-turbulence interaction. Standard methods to model turbulence like Reynolds-Averaged Navier-Stokes equations (RANS) and hybrid RANS-Large Eddy Simulations (LES) models are unable to reproduce the local turbulence dynamics observed in cavitating flow experiments. In an effort to facilitate the development of accurate RANS modelling procedures for cavitating flows, a data-driven approach devised by integrating Gene-Expression Programming (GEP) into a traditional RANS approach, without continuous RANS model feedback is proposed. The GEP algorithm is employed to obtain an additional corrective term, composed of flow variables derived directly from a baseline URANS cavitating flow calculation to be fit directly in the Boussinesq approximation. In addition, an optimizer based on the BFGS algorithm is employed on top of the GEP algorithm to ensure the solution is independent of the uncertainties associated with GEP. To evaluate the complexity of the model, we compare the approach with a more elementary linear regression technique. The GEP-CFD method is then used to train two expressions, for the Reynolds shear stress and the Turbulent Kinetic Energy simultaneously. The GEP-CFD approach is able to produce the flow features not captured by baseline RANS simulations in several areas, outperforming the RANS simulations and thus projects as a promising approach to augment numerical simulations for cavitating flows.
\end{abstract}



\begin{keyword}
Cloud Cavitation \sep Turbulence Modelling \sep Gene-Expression Programming \sep Machine Learning 

\end{keyword}

\end{frontmatter}

%\printnomenclature
\textbf{Nomenclature}

{\renewcommand\arraystretch{1.0}
\noindent\begin{longtable*}{@{}l @{\quad=\quad} l@{}}
$u_{i}$  & velocity component in the ith direction \\
$u_{j}$ &  velocity component in the jth direction \\
$\rho_{m}$& density of the mixture phase \\
$\mu_{m}$ & viscosity of the mixture phase \\
$\vec{u}$ & velocity field \\
$p$   & pressure field \\
$\rho_{l}$ & liquid density \\
$\rho_{v}$ & vapor density \\
$\mu_{l}$  & liquid dynamic viscosity \\
$\mu_{v}$ & vapor dynamic viscosity \\
$\mu_{t}$ & turbulent eddy viscosity \\
$\alpha_{l}$ & liquid void fraction \\
$\alpha$ & vapor void fraction  \\
$m^{+}$  & evaporation source term \\
$m^{-}$  & destruction source term \\
$p_{sat}$  & saturation pressure \\
$t_{\infty}$  & free stream time scale \\
$U_{\infty}$  & free stream velocity \\
$C_{dest}$  & Vapor destruction constant \\
$C_{prod}$  & Vapor production constant \\
$\tau_{ij}$ & Reynolds stress tensor \\
$S_{ij}$ & mean strain rate tensor \\
$\delta_{ij}$ & Kronecker delta function \\
$k$ & Turbulent Kinetic Energy \\
$n_{a}$ & Maximum number of arguments for operator \\
$x$ & Reynolds stress ($\tau_{12}$) \\
$\alpha$ & void fraction \\
q & Time-Averaged standard deviation of Velocity in Stream-wise direction \\
r & Time-Averaged standard deviation of Velocity in wall direction \\
u & time-averaged velocity in direction of flow \\
v & time-averaged velocity in direction to perpendicular to the flow \\
$C_{efgh}$ & Constants for linear equation in LR \\
$N$ & degree of LR polynomial \\
$\textbf{z}_{i}$ & Feature vector \\
\textbf{Z} & Feature matrix \\
\textbf{c} & Coefficient vector \\
\textbf{y} & target vector \\
$\gamma$ & LR regularization parameter \\
\textbf{I} & Identity matrix \\
$l_{i}$ & optimized coefficient \\
$I_{field}$ & cumulative influence of each field \\
\textit{t} & component in polynomial expression \\
$x_{j}$ & field of interest \\
$x^{i}_{-j}$ & fields except the field of interest \\
T & total number of trees for random tree analysis \\
nodes(t) & set of all nodes in tree \textit{t} \\
I(param) & indicator function \\
$f_{n,ij}$ & closure expression for \textit{u'v'} \\
$b_{n,ij}$ & closure expression for TKE \\
\end{longtable*}}
\section{Introduction} \label{introduction}
Cavitation is defined by the formation of vapor cavities when the local pressure drops drastically below the saturated vapor pressure. It is a complex, multi-phase and multi-scale flow phenomenon applicable in a multitude of engineering applications. It poses as a detrimental occurrence in hydraulic machinery like marine propeller, water turbines and underwater vehicles as it leads to noise, vibration and erosion damage responsible for reducing the machine's efficiency \cite{brennen2005fundamentals}. Cavitation can also function as a beneficial tool like drilling for geothermal reservoirs and bio-diesel production \cite{ji2006preparation}. Therefore, it is essential to investigate cavitating flows and control the underlying dynamics to reduce or employ cavitation. Multiple studies have studied cavitation using experimental techniques and numerical methods. \\
In the field of experimental methods, Knapp \cite{knapp1955recent} was among the first ones to study cavity dynamics using high-speed motion pictures five decades ago. He observed the presence of a re-entrant jet moving in a reverse direction rather than the direction of the flow and breaking off the main vapor cavity. He established the presence of the re-entrant jet was responsible for the periodic shedding of the cloud cavity. More recently, there have been several studies to study the mechanisms driving cavitation \cite{arndt2000instability, ganesh2016bubbly, stutz2003x, ge2022intensity}. These studies have indicated that apart from the re-entrant jet, the presence of condensation shock emanating from the collapse of the vapor cavities is also responsible for the periodic shedding. These works have been followed with numerical works to validate the experiments and propose accurate numerical models for cavitation. Numerical studies often involve the coupling of a cavitation model and a turbulence model to simulate the cavitation-turbulence interplay. Indeed, turbulence models play a pivotal role in this coupling, with the statistically averaged flowfields garnering significant interest and offering valuable insights into the flow. A significant number of such works often use the Reynolds-Averaged Navier-Stokes (RANS) models or the hybrid RANS-Large Eddy Simulation (LES) models due to their low computational cost. \\
However, these approaches are unable to predict the local turbulence dynamics accurately and display considerable discrepancies \cite{coutier2003evaluation, apte2023numerical}. These discrepancies can be largely accounted to the use of a linear-stress relation, the Boussinesq approximation \cite{leschziner2015statistical} to calculate the Reynolds stress and close the RANS equations. While computing high-fidelity simulations like a LES could provide a viable alternative \cite{bhatt2020numerical} , the high computational cost still pose a challenge, therefore making RANS a preferred tool to simulate cavitating flow simulations. Hence, more accurate approximations of the Reynolds stress term are crucial in ensuring accurate RANS simulations. 
%Therefore, there exists a need to bridge low-fidelity simulation techniques with high-fidelity simulation/ experimental data by employing machine learning techniques. These techniques, motivated by high-fidelity datasets could drive the low-fidelity computational models to reduce the latter's discrepancies. 

The rise of data-driven techniques have provided inspiration for augmenting turbulence modelling as well. More and more data is generated from high-fidelity simulations and experiments with more advanced equipment. This high-fidelity data can be leveraged with machine learning (ML) techniques to augment RANS modelling. While data-driven techniques like ML and deep learning have been utilized to investigate a myriad number of turbulent flows \cite{duraisamy2019turbulence, zhang2019recent}, few studies have been employed data-driven methods for cavitating flows.  Xu et al. \cite{xu2021rans} used a random forest machine-learning technique, fed with high-fidelity LES datasets to simulate flow around a Clark-Y hydrofoil. They observed the method was able to capture the shedding process of the cloud cavity. A similar technique was employed by Sikirica et al. \cite{sikirica2020cavitation} to predict the empirical constants in a mass transfer model that simulates the vapor generation and destruction processes. However, the stochastic nature of random forests itself yields considerable uncertainty and a significant amount of training data is required to make such a universal constitutive relation. Zhang et al. \cite{zhang2023data} employed a neural network to train and predict the linear and non-linear parts of the anisotropic Reynolds stress tensor. Initially, the velocity divergence term is input from the phase transition-rate to accommodate cavity dynamics followed by an implicit and explicit neural network loops to improve the computational stability. The model was trained using LES data for two cases: cavitating flow in a venturi and cavitating flow over a hydrofoil. They observed the data-driven model was much more accurate than the URANS calculation in predicting the unsteady development of the cavity and the mean velocity. Another approach, proposed by Gao et al. \cite{gao2024towards} used a graph neural network to model cavitating flow in laminar flow conditions. Their approach used a hypergraph connecting nodes by elements, similar to finite element method to predict the evolution of primary cloud cavity over a hydrofoil. There have been some studies coupling numerical method with deep learning for cavitation in marine engineering devices \cite{li2024deep, choi2023composite}. However, generating a training dataset for neural networks and other such techniques to model cavitating flows is prohibitively expensive in terms of computational cost and thus, poses a significant challenge.

A leading goal of data-driven turbulence modelling is the inclusion of embedded variance based on the mean flow quantities for non-linear closure turbulence models, based on Pope's viscosity assumptions \cite{pope1975more}. Amongst the various training approaches, symbolic regression techniques have a particular advantage, as they generate easy-to-implement and interpretable closed-form mathematical equations. Such a  novel data-driven technique was introduced by Weatheritt \& Sandberg \cite{weatheritt2016novel}. Titled Gene-Expression Programming (GEP) and inspired by the works of Ferreira \cite{andida2001gene}, this method generated Galilean invariant explicit algebraic Reynolds stress model (EARSM) closures to improve the stress-strain relation, replacing the Boussinesq approximation. The method has been applied to model various turbulent flows applications. Akolekar et al. \cite{akolekar2019development} developed closure relations for modelling wake mixing using GEP. Schoepplein et al. \cite{schoepplein2018application} used GEP to model the subgrid scale term for LES simulations to model the turbulent transport in pre-mixed flames. Hammond et al. \cite{hammond2022machine} employed the GEP algorithm to learn nonlinear expressions for the Reynolds stresses, in the context of designing internal cooling channels in turbine blades. Liu et al. \cite{liu2024priori} explored the use of GEP to simultaneously develop the sub-grid scale stress and heat-flux models for LES simulations of natural convection. However, GEP has several algorithm parameters that control the model closure expression, leading to an increased dependence on the input data and parameters.   

In summary, the generation of a training dataset poses a significant computational challenge for employing neural networks or ML algorithms to augment turbulence modelling of cavitating flows. While GEP may provide a viable alternative through its requirement of a comparatively smaller dataset, the parameter-dependence on the model closure expression continues to be a source of uncertainty. Thus, this study aims to bridge the two gaps by employing a GEP approach to model cavitating flows and then utilizing an optimization technique atop the GEP algorithm to reduce the parameter uncertainties and provide a corrective term that is able to capture the cavitation-turbulence interplay dynamics observed in experiments. To our knowledge, there have been no previous attempts to apply a symbolic regression method like GEP to develop turbulence closures for accurately modelling cavitating flows. Therefore, this study demonstrates a broader impact of ML for a multi-phase flow like cavitation.

This paper is organized as follows: Section \ref{se:pap3_methodology} discusses the proposed methodology outlining the numerical models, followed by an introduction to GEP and the linear regression routine. Next, the case setup is described. Finally, Section \ref{se:pap3_result} discusses the performance of the GEP-CFD approach while Section \ref{se:pap3_conclusions} provides conclusions and future scope of the existing study.

\section{Methodology}\label{se:pap3_methodology}
\subsection{Numerical Methods}\label{ss:pap3_nm}

Numerical simulations designed to model cavitating flows involve the coupling of cavitation model and a turbulence model: while there exists a huge array of cavitation models \cite{folden2023classification}, Transport-Equation Models (TEMs) have been especially gaining popularity over the years. These models consider both the liquid and vapor phases as strongly coupled and define vapor formation (evaporation) and vapor destruction (condensation) terms as one quantity in the governing equations while defining them differently.  
\begin{eqnarray}
\frac{\partial (\rho_{m} u_{i})}{\partial t}+\frac{\partial (\rho_{m} u_{i} u_{j})}{\partial x_{j}}\nonumber\\
=-\frac{\partial p}{\partial x_{i}}+\frac{\partial}{\partial x_{j}}[(\mu_{m} +\mu_{t})(\frac{\partial u_{i}}{\partial x_{j}}+\frac{\partial u_{j}}{\partial x_{i}}- \frac{2}{3}\frac{\partial u_{k}}{\partial x_{k}} \delta_{ij})]
\end{eqnarray}

\begin{equation}
    \frac{\partial\rho_{l} \alpha_{l}}{\partial t}+\frac{\partial(\rho_{l} \alpha_{l} u_{j})}{\partial x_{j}}= \dot{m}^{-} + \dot{m}^{+}
\end{equation}
\begin{equation}
    \rho_{m}= \rho_{l}\alpha_{l}+\rho_{v}\alpha
    \label{eq:mixture_density}
\end{equation}
\begin{equation}
    \mu_{m}= \mu_{l}\alpha_{l}+\mu_{v}\alpha
\end{equation}
where $u_{i}$ and $u_{j}$ are the velocity components in the ith and jth directions respectively,  $\rho_{m}$ and $\mu_{m}$ are respectively the density and viscosity of the mixture phase, \textit{u} is the velocity, \textit{p} is the pressure, $\rho_{l}$ and $\rho_{v}$ are respectively the liquid and vapor density, $\mu_{l}$ and $\mu_{v}$ are respectively the liquid and vapor dynamic viscosity while $\mu_{t}$ represents the turbulent viscosity. $\alpha_{l}$ and $\alpha$ are respectively the liquid and vapor void fraction defined as $\alpha_{l} + \alpha = 1$. The source ($\dot{m}^{+}$) and sink ($\dot{m}^{-}$) terms represent the condensation (vapor destruction) and evaporation (vapor formation) terms respectively. Over the years, several Transport-Equation Models (TEM) have been defined \cite{merkle1998computational, zwart2004two}. 

The Merkle computational model \citep{merkle1998computational} has been used throughout this work for the cavitation model. This model is derived for a cluster or a cloud of bubbles. Unlike other Transport-Equation Models, the Merkle model is not derived from the Rayleigh-Plesset equation thus averting the equation's shortcomings resulting from its assumptions. The mass fraction form of the condensation and destruction terms is defined as:
\begin{equation}
    \dot{m}^{-}= \frac{C_{dest} min(p-p_{sat},0) (1-\alpha) \rho_{l}}{0.5 U_{\infty}^{2} t_{\infty}\rho_{v}}
\end{equation}
\begin{equation}
    \dot{m}^{+}= \frac{C_{prod} max(p-p_{sat},0) \alpha} {0.5 U_{\infty}^{2} t_{\infty}}
\end{equation}
where $\rho_{v}$ and $\rho_{l}$ are the vapor density and the liquid density respectively, $p$ and $p_{sat}$ are the pressure and the saturation pressure respectively, $t_{\infty}$ is the free stream time scale and $U_{\infty}$ is the free stream velocity.  The empirical factors $C_{dest}$ and $C_{prod}$ are set as 80 and $10^{-3}$ respectively. These values are the default values used in the solver.

Regarding the turbulence model, the standard k-$\omega$ Shear Stress Transport (SST) \cite{menter2003ten} is utilized in this study. The model uses the Boussinesq approximation, a linear constitutive relationship to formulate the Reynolds stress anisotropy defined as:
\begin{equation}
    \tau_{ij} = 2 \mu_{t} S_{ij} - \frac{2}{3} k \delta_{ij}
\end{equation}
where $\tau_{ij}$ is the Reynolds stress tensor, $S_{ij}$ is the mean strain rate tensor, $\delta_{ij}$ is the Kronecker delta function and \textit{k} is the turbulent kinetic energy. $S_{ij}$ and $\delta_{ij}$ are defined as follows:
\begin{equation}
    S_{ij} = \frac{1}{2}(\frac{\partial \bar{u_{i}}}{\partial x_{j}}+\frac{\partial \bar{u_{j}}}{\partial x_{i}})
\end{equation}
\begin{equation}
    \delta_{ij} = \begin{cases}
    1,& \text{if } i = j\\
    0,              & \text{otherwise}
\end{cases}%% 
\end{equation}

The motivation to employ these standard models for the simulation is to evaluate if the data-driven techniques are able to perform on the same level of accuracy, if not superior to standard modelling techniques.

\subsection{Gene-Expression Programming}\label{ss:gep}

\begin{figure}
    \centering  
    {\includegraphics[width=15cm, height=8cm]{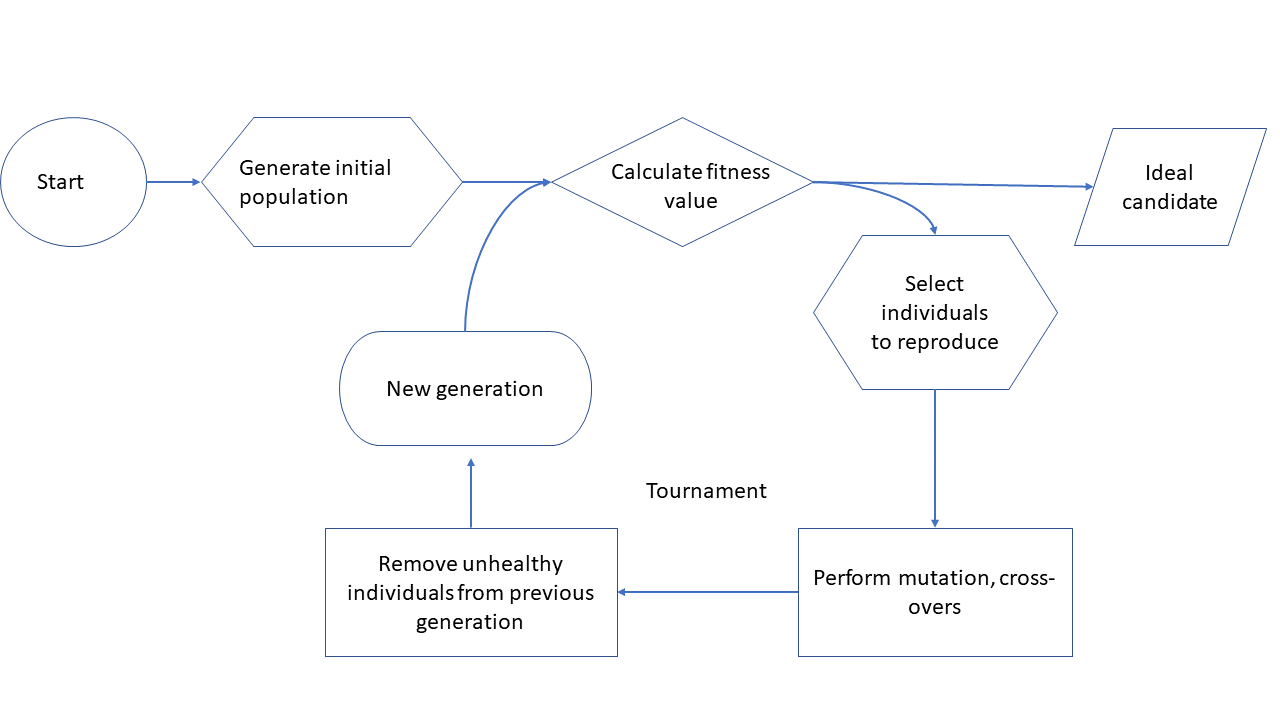}};      
  \caption{Flowchart outlining the GEP process}
  \label{gep_flowchart}
\end{figure}

GEP, being an Evolutionary Algorithm (EA) is driven by the Darwinian principle of nature, "Survival of the Fittest". The procedure employs an in-house tool, initially developed and tested in \cite{weatheritt2016novel}. Before discussing the methodology of the GEP, the fitness function \textit{F} is defined as the deviation between the training data and the model. The objective is to find a function \textit{f = f(x,y)} where \textit{n} set of data-points are provided (\textit{$f_{k},x_{k},y_{k}$}) as the training data. The general aim of the algorithm is to search the space of all functions for the one that fits the 'best' the training dataset. The fitness function of a candidate solution $f^{guess}$ is defined as :
\begin{equation}
  F = 1 - \frac{1}{n} \sum_{k=1}^{n} \frac{\lvert f^{\text{guess}}(x_{k},y_{k}) - f_{k} \rvert}{f_{k}}
\end{equation}
thus, utilizing the relative distance from each data point. The optimal fitness will be when $f^{guess} = f_{k}$ or, 1.
Fig \ref{gep_flowchart} provides insights into the processes driving GEP. The GEP first starts to create an initial population of a pre-defined size. Then, it selects the healthiest individuals based on their fitness values. Following the selection, the algorithm then uses a tournament procedure to reproduce the offspring of healthy individuals. Each tournament consists of a specific number of candidates and only a limited number are allowed to advance to the next generation. The losing candidates are replaced by the offspring of the successful candidates at the next generation. Reproduction happens due to cross-over and mutation procedures. Thus, an optimal candidate is obtained when either
\begin{enumerate}
    \item The convergence criterion is achieved or,
    \item A prescribed number of generations has been completed
\end{enumerate} 
In addition, all candidate solutions, especially the ideal candidate one, are governed by their length: a chromosome's string for each gene is divided into two parts, the head and the tail. The head length is fixed, while the tail length is determined by the head length. For example, if a candidate has a head length $h$, then its tail length will be $t = h \cdot (n_{a} + 1)$, where $n_a$ is the maximum number of arguments for a mathematical operator ($\times$, $\textminus$, etc.) or its arity. For instance, the mathematical operator $\times$ has an arity of two ($a \times b$), while the operator $\sin$ has an arity of one ($\sin(x)$).

To utilize GEP and bridge the discrepancies between experiments and CFD simulations, a corrective function is defined for the term $f_{ij}$ in the Boussinesq approximation. The term is a linear combination of flow parameters influencing the cavitating flows and dimensional constants. The term is described as:
\begin{equation}
    f_{ij} = f(x,u,v,\alpha)
\end{equation}
where \textit{x} is the Reynolds stress ($\tau_{12}$) from the CFD simulations, a direct output from the Boussinesq approximation, $\alpha$ is the void fraction and \textbf{u,v} are the time-averaged velocities in the direction of the flow and direction perpendicular to the flow respectively. The objective behind using $\alpha$ is to ensure the GEP is able to capture the cavity dynamics that inherently influence the turbulence fields \cite{apte2023numerical} and \textbf{u,v} to ensure the constitutive relation is independent of the device geometry. The \textit{u'v'} term is selected at the onset to formulate the GEP equation. This technique is called frozen training as the aim is to define a closure expression against a high-fidelity database based on the candidate expression's fitness function. It needs to be highlighted that in frozen training, the model uses fixed inputs from the supplied training data to derive the symbolic closure expressions that are subsequently compared to the same fixed database. The frozen training approach therefore does not account for potential changes of input data occurring in actual CFD runs. To provide some context, the fitness value of a baseline k-$\omega$ SST model \cite{menter2003ten} calculation is 1.93887. 

Indeed, the idea behind sole use of \textit{u'v'} is theoretically infeasible: as the extra anisotropic terms are added to the Boussinesq hypothesis, these terms will influence other Reynolds stresses. Therefore, these anisotropic terms will be later introduced in the turbulent kinetic energy term as well to maintain the constitutive relations regardless of the Reynolds stress term.  

\subsection{Linear Regression routine}\label{ss:lr}

To compare the efficiency of the GEP-assisted framework, a linear regression procedure is performed. A linear regression algorithm aims to model the relationship between the target variable (\textit{u'v'} obtained from experiments, in this case) and component variables by fitting a linear equation.
\begin{equation}
    \textit{u'v'}_{exp} = P(x, u, v, \alpha) = \sum_{e+f+g+h \leq N} c_{efgh} x^{e} u^{f} v^{g} \alpha^{h}
\end{equation}

where $c_{efgh}$ are the coefficients associated with the basis polynomials. \textit{N} is the degree of the polynomial and is a user-input function but has to be conserved to avoid ill-conditioning of the predictor matrix and numerical instability. For each data point, a feature vector \textbf{z}$_{i}$ is defined 
\begin{equation}
    \mathbf{z_{i}} = [1, x_{i}, u_{i}, v_{i}, \alpha_{i}, x_{i}^{2}, x_{i}u_{i}, \ldots, x_{i}^{e}]
\end{equation}

Stacking the feature vectors \textbf{z}$_{i}$ for all data points results in a feature matrix \textbf{Z}. Similarly, a coefficient vector \textbf{c}, consisting of all the coefficients and target vector \textbf{y} containing the target values for each data point can be constructed. The coefficients are determined by standard Ordinary Least Squares (OLS) matrix, or minimizing the residual sum of the squares:
\begin{equation}
    \textbf{c} = (\textbf{Z}^{T} \textbf{Z})^{-1} \textbf{Z}^{T} \textbf{y}
\end{equation}
To evaluate the accuracy of the linear equation with the training data, the residual sum of squares (RSS) is calculated:
\begin{equation}
    RSS = ||\textbf{y} - \textbf{Z} \textbf{c}||^{2} 
\end{equation}
An ideal RSS will occur when the feature vector and coefficient vector are able to perfectly predict the target vector, $\textbf{y} =\textbf{Z} \textbf{c}$, thus obtaining a value of zero.  
To prevent the issues related to \textit{N} mentioned above along with data over-fitting, a penalty term of the second order is added to the loss function, known as the Ridge regression or the L2 regularization:
\begin{equation}
    RSS = ||\textbf{y} - \textbf{Z} \textbf{c}||^{2} + \lambda ||\textbf{y}||^{2}
\end{equation}
where $\lambda$ is a regularization parameter controlling the strength of the penalty term. The term $||c||^{2}$ is the L2 norm of the coefficient vector:
\begin{equation}
    ||c||^{2} = \textbf{c}^{T} \textbf{c}
\end{equation}
Thus, the coefficients are calculated as:
\begin{equation}
    \textbf{c} = (\textbf{Z}^{T} \textbf{Z} + \lambda \textbf{I})^{-1} \textbf{Z}^{T} \textbf{y}
\end{equation}
where \textbf{I} is the identity matrix. 
The current model takes in the four variables and obtains a polynomial expression for a 7th degree polynomial, thus setting $N=7$. The fundamental difference between the GEP framework and the linear regression routine is that the GEP framework takes in the variables and derives the constitutive relation based on user input; this indicates some physics could be used to inform the framework in contrast to the linear regression, which obtains the relation solely on the basis of fitting as many data points as possible.   

\subsection{Case Setup}\label{ss:case_setup}

\begin{figure*}
\includegraphics[width=17cm, height=3cm]{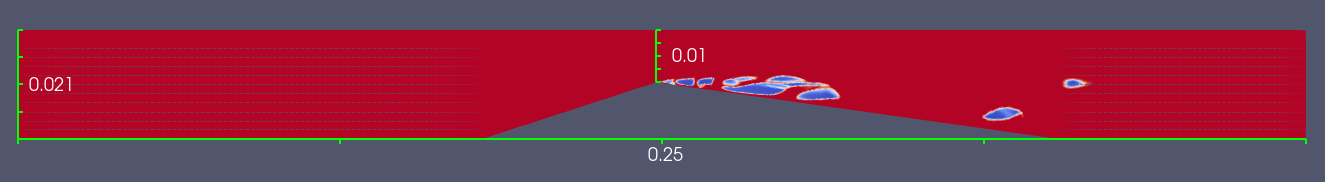}
\caption{The venturi-type geometry. Dimensions in mm. Here the vapor bubble regions are denoted by blue color while the red region denotes water. Flow direction is from left to right}
\label{venturi2}
\end{figure*}

A converging-diverging nozzle (venturi) is used as the case geometry. As seen in Fig \ref{venturi2}, the venturi has an 18\degree convergent angle and 8\degree divergent angle with cavitation inception at the throat. The height of the venturi channel is 21 mm at the inlet and reduces to 10 mm at the throat. The experimental data was taken from high-speed Particle-Image Velocimetry experiments described in \cite{ge2021cavitation}. It was observed that the mean cavity length reproduced in the CFD simulations was identical to the one measured in the experiments.

\section{Results}\label{se:pap3_result}
\subsection{Scalar Regression for \textit{u'v'}}\label{ss:uv}
\begin{figure}[htbp]
    \centering
    \includegraphics[width=1.5\linewidth]{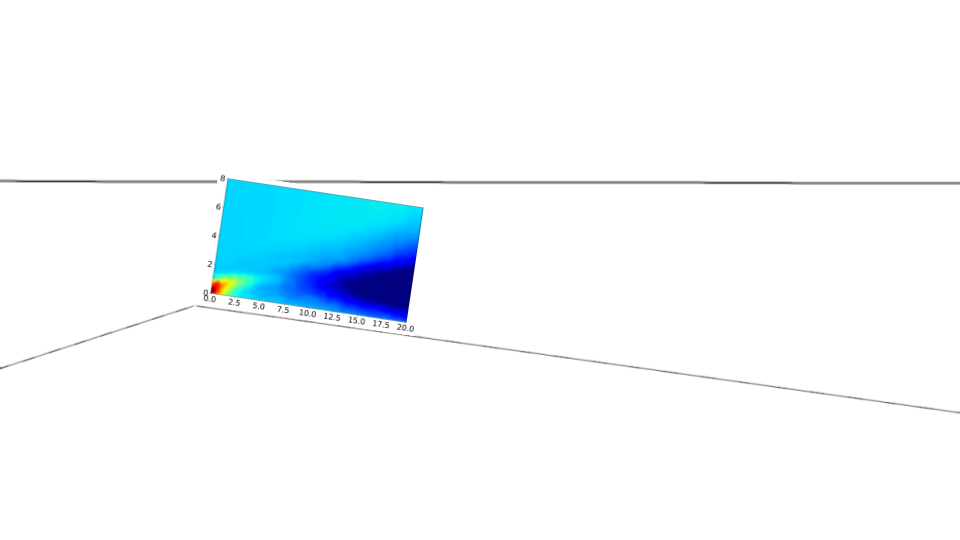}
    \caption{Reynolds shear (\textit{u'v'}) from the experimental data \cite{ge2021cavitation}. For ease of comparison, the axes will be tilted throughout the remaining study}
\end{figure}
\begin{figure}[htbp]
    \centering
    \includegraphics[width=0.45\linewidth]{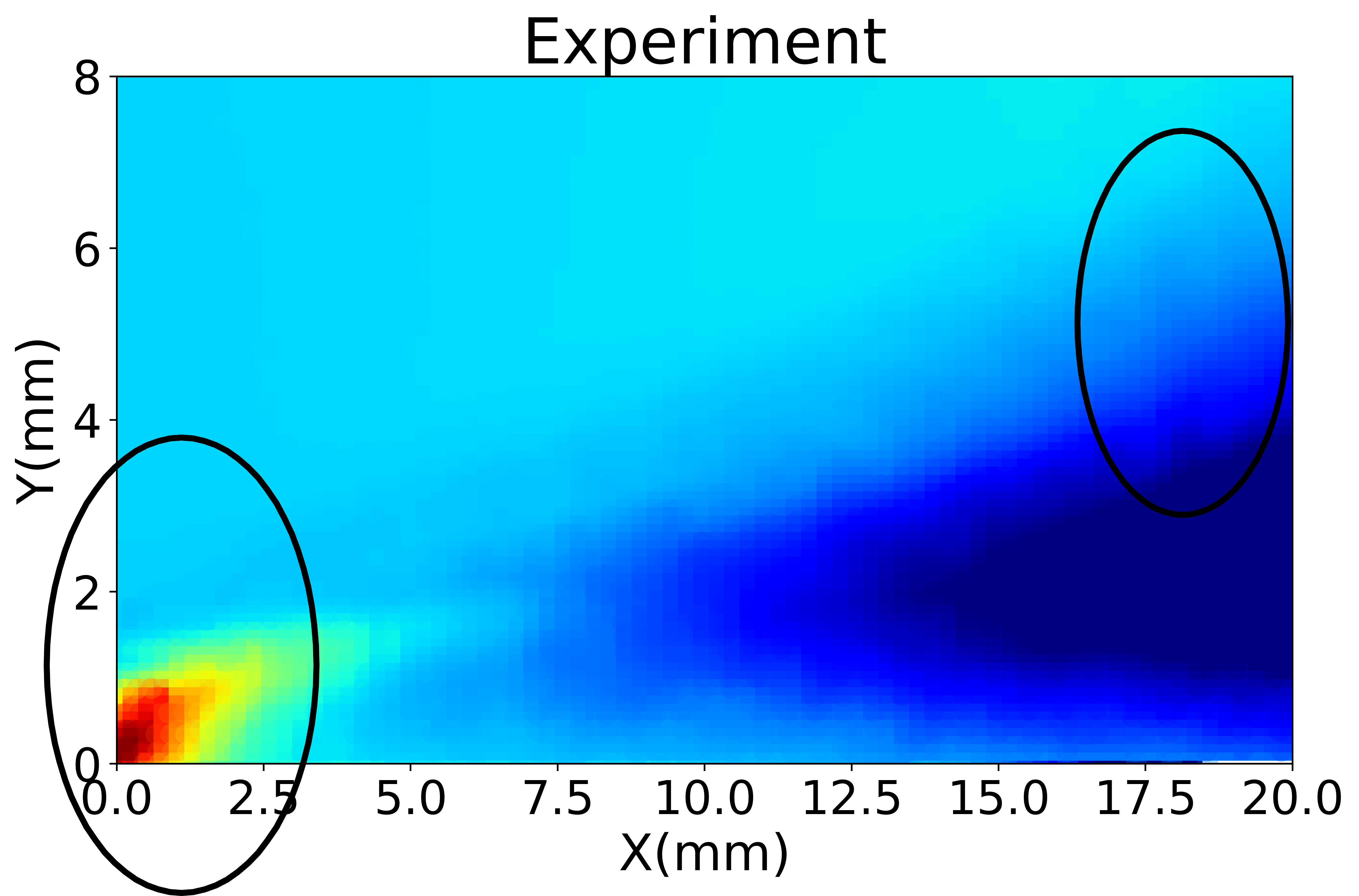}
    \includegraphics[width=0.45\linewidth]{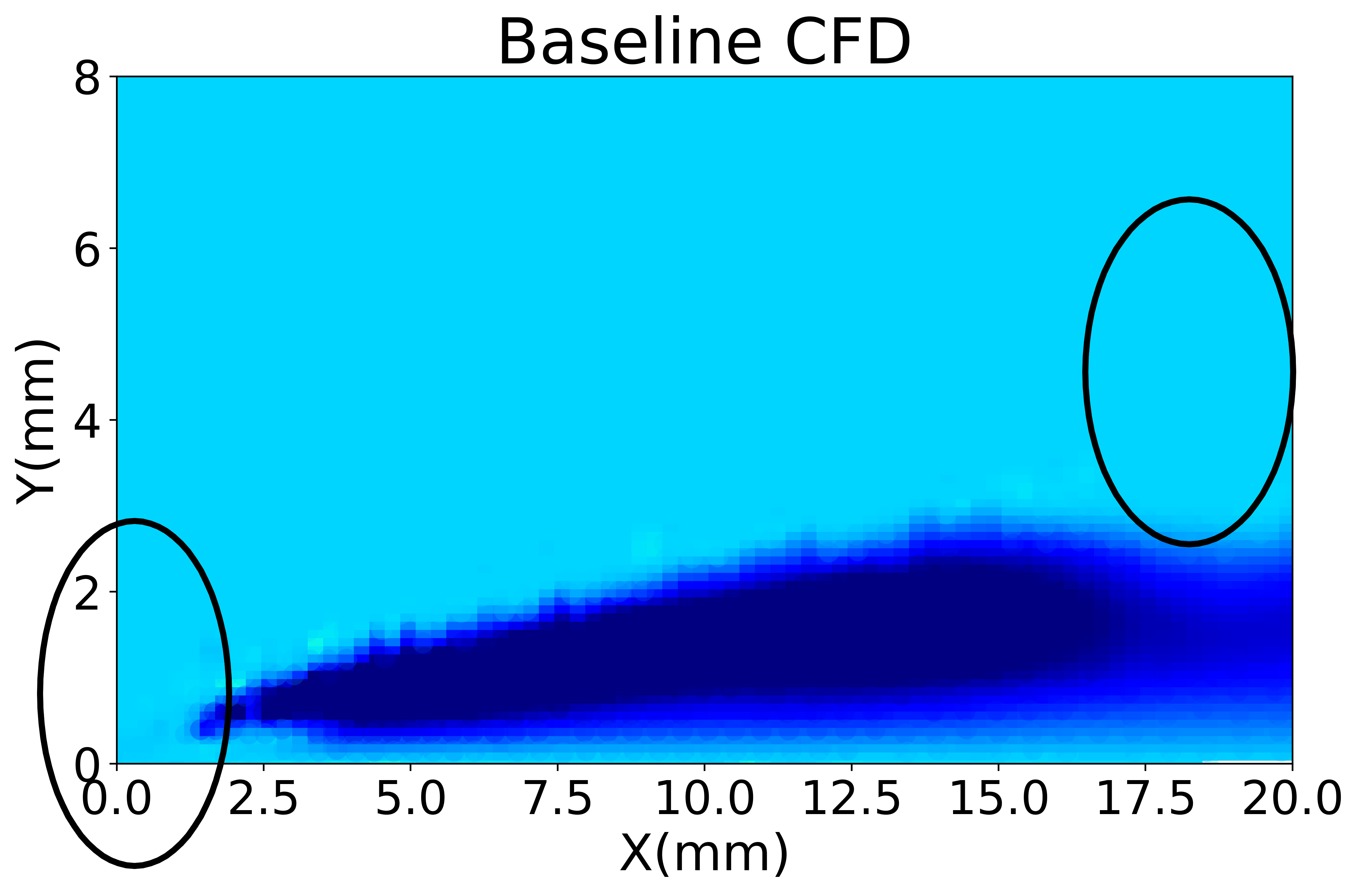}
\end{figure}
\begin{figure}[htbp]
    \centering
    \includegraphics[width=0.9\textwidth]{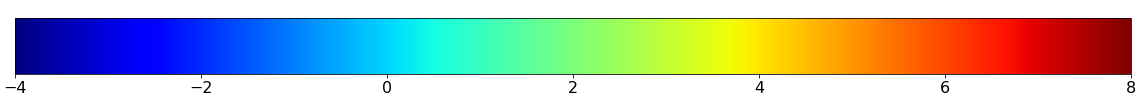}
    \caption{Reynolds shear stress \textit{u'v'} for both a) experiments and b) CFD simulation with the origin being the venturi's throat. The circles denote that the CFD simulation is unable to reproduce the high stress values close to the throat and the high \textit{u'v'} values downstream away from the wall}
    \label{uv_gep}
\end{figure}
Fig \ref{uv_gep} (a) and (b) represent the Reynolds shear stress \textit{u'v'} from the experiments and CFD respectively. Both these figures have been plotted with the throat of the venturi taken as the origin. While the CFD results generated from the Boussinesq approximation are able to predict the mean cavity shape, they are unable to capture high Reynolds stress observed near the throat where the primary cavity develops and collapses and downstream where the detached cavity is expected to roll-up and collapse as it exits the low-pressure region. We now proceed to apply the GEP procedure on the given data.

To implement GEP, we use a framework titled "EVE" or EVolutionary algorithm for the development of Expressions \cite{weatheritt2016novel} that symbolically regresses tangible algebraic equations from the training data (\textit{$u'v'_{truth}$,x,u,v}, $\alpha$), defined as expressions. 
\begin{figure}[htbp]
    \centering
    \includegraphics[width=\linewidth]{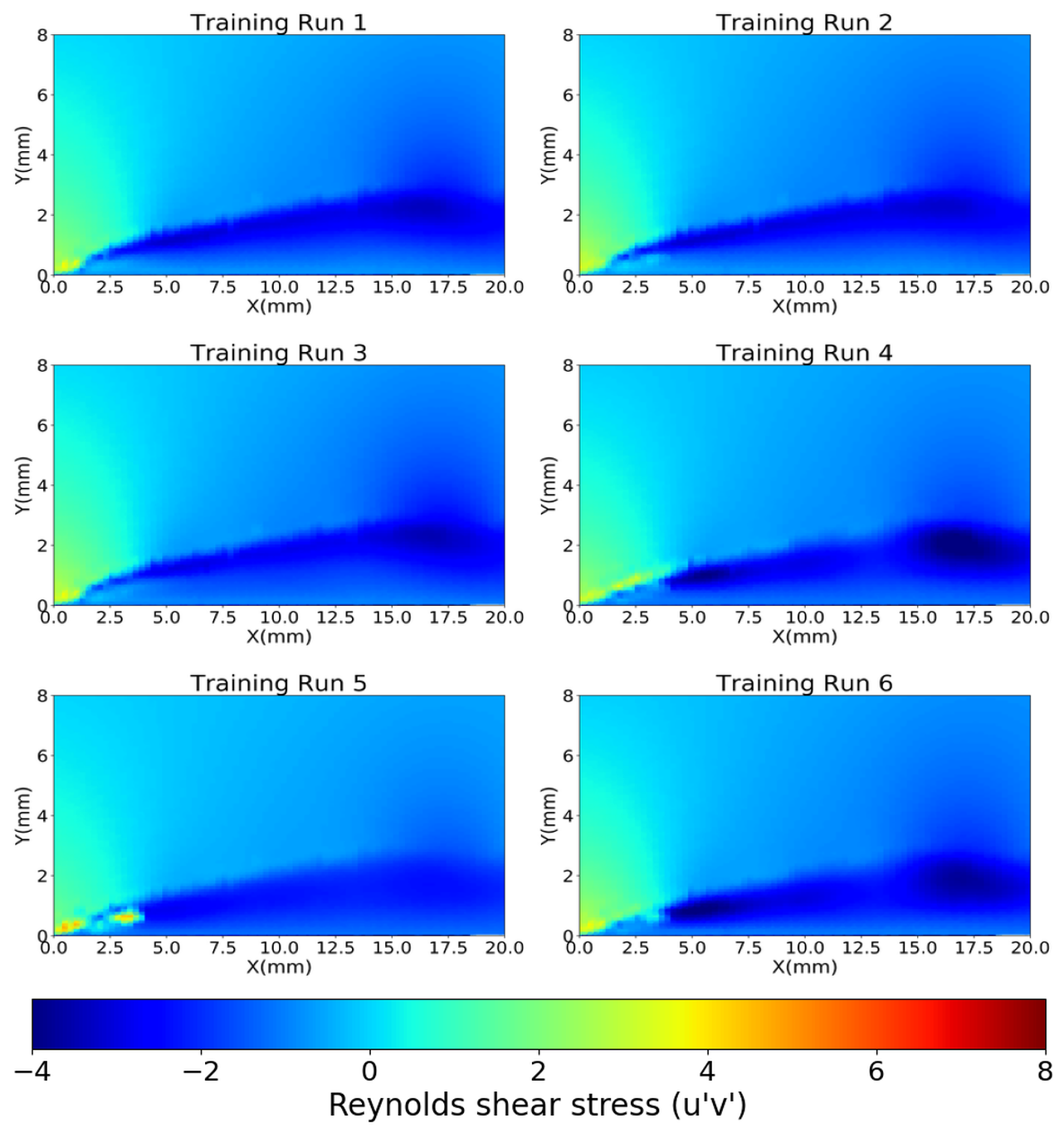}
    \caption{Scatter plots for Re stress from five formulations provided from six different GEP training runs with the origin being the venturi throat. All plots predict turbulence dynamics identical to CFD data}
  \label{fig:five_gep}
\end{figure}

Initially, six GEP training runs are run consecutively. These training runs are conducted in a near-identical manner, when one parameter in the GEP procedure (shown in Table \ref{list:params}) is changed, generating distinct expressions. These expressions are:

\begin{eqnarray}
\hspace*{-2cm} f_{1,ij} &=& 0.21(-3\alpha + 1v + 1x + 0.32)(-1\alpha + 0.1u - 0.1v + 0.1x + 2.4)  \label{Eq:19}\\
\hspace*{-2cm} f_{2,ij} &=& 0.21(-3\alpha + 1v + 1x + 0.19)(1\alpha + 0.1u + 0.1x + 0.32) \label{Eq:20} \\
\hspace*{-2cm} f_{3,ij} &=& 0.21(-3\alpha + 1v + 1x + 0.33)(0.1u - 0.1v + 0.2x + 1.79) \label{Eq:21} \\
\hspace*{-2cm} f_{4,ij} &=& -1\alpha - (0.02u + 0.18)(1\alpha + 1.0vx - 1.0v - 1x)\label{Eq:22} \\
\hspace*{-2cm} f_{5,ij} &=& -0.15\alpha - 0.43v(1\alpha - 2.0) - [1\alpha - (1\alpha - 1.0)(1x - 1.0)](1\alpha - 0.21x + 0.1) + 0.05 \label{Eq:23} \\
\hspace*{-2cm} f_{6,ij} &=& -1\alpha + (-2.0\alpha + 1v + 0.86)(0.03u - 0.28x + 0.06) \label{Eq:24}
\end{eqnarray}

\begin{table}
    \centering
    \begin{tabular}{cc}
        Parameter Name & Abbreviation\\
        Number of Operators & 3\\
        Number of Random Constants & num\textunderscore rnc\\
        Domain of Random Constants & [integer, integer]\\
        Number of Constants & 2\\
        Probability of Occurrence of Variable or Constant & prob\textunderscore ar2 \textunderscore symbols\textunderscore  \textit{arity}\\
        Specific Probability of Individual Variables & aprob \textunderscore ar\textunderscore\textit{arity}\\
        Head Length & head\textunderscore length\\
        Number of Genes & ngenes \\
        Size of Initial Population & pop\textunderscore size\\
        Number of Generations & ngens\\
        Number of Colonies competing in a single Tournament & tsize\\
        Number of Colonies that advance to next Generation & msize\\
        Probability of Genes Mutating or performing Cross-overs & p\textunderscore \textit{arity}\\ 
    \end{tabular}
    \caption{List of parameters in the GEP algorithm along with their abbreviation. Changing one parameter yielded a different expression for \textit{u'v'}. Here, \textit{arity} is defined by the number of input arguments a symbol takes.}
    \label{list:params}
\end{table}
All the expressions have the polynomial degree 2 with the exception of Eq. \ref{Eq:23}, having a degree of 3. While all these expressions have been generated from separate training runs, they show a considerable amount of similarity with each other especially, Eqns. \ref{Eq:19}-\ref{Eq:21}. All the trained expressions show the influence of all the functions u,v,x,$\alpha$ for the target \textit{y}, specially the void fraction, $\alpha$ which highlights the role of cavitation in the turbulence. The expressions resulting from these different training runs are then used to plot the Reynolds stress as shown in Fig \ref{fig:five_gep}. All the equations are simulating the Reynolds shear stress field identical to CFD data but are not able to reproduce the discrepancies observed between the CFD and experimental plots in Fig \ref{uv_gep}. In other words, they are performing as accurately as the baseline CFD simulations. In addition, the sizeable number of parameters, listed above in \ref{list:params}, result in a considerable number of uncertainties. Indeed, since GEP is non-deterministic, the change in a single parameter does result in a distinct expression. But it would be exhausting to run countless GEP training runs to evaluate the fittest expression. 
\begin{figure}
    \centering  
    {\includegraphics[width=15cm, height=8cm]{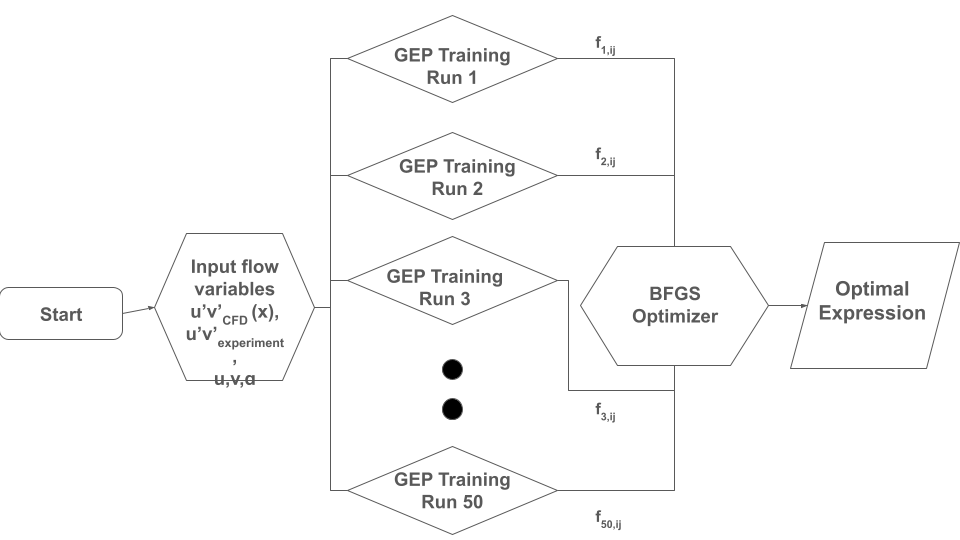}};      
  \caption{Flowchart outlining the optimization process. The flow variables are input into GEP training runs, where each run is caused by one parameter change. These expressions, in the form of $a_{n, ij}$ are then input into a BFGS optimizer to yield a combined optimal expression. }
  \label{opt_flowchart}
\end{figure}
In order to make the framework independent of the uncertainties associated with GEP's parameters, an optimizing technique is applied atop the GEP algorithm, as shown in Fig \ref{opt_flowchart}. An ensemble set of 50 optimal different training expressions ($f_{i}$) are selected. These expressions have a fitness value close to 1, thus indicating their proximity to the training data and are formed as a result of the GEP parameter changes. To define a optimal expression for the Reynolds stress, these expressions are optimized for a constitutive relation such that $f_{optimal} = \sum_{i=1}^{50} l_{i} f_{i}$ where $l_{i}$ is the optimized coefficient of the expression $f_{i}$. These basis functions are obtained by minimizing the error function associated with each expression. The error is minimized by calculating the Hessian, the matrix composed of second-order derivatives of the system variables, used to determine the minima of the error function. The approach employs the Broyden-Fletcher-Goldfarb-Shanno algorithm \cite{broyden1970convergence, fletcher1970new, goldfarb1970family, shanno1970conditioning}, a quasi-Newton method that uses an approximate Hessian rather than compute the exact Hessian matrix to reduce computational time.  The optimized result is able to determine the accuracy of each expression with respect to the truth and determine a basis function that elevates or downgrades the contribution of each expression to the final expression.

\begin{figure}[htbp]
    \centering
    \includegraphics[width=\linewidth]{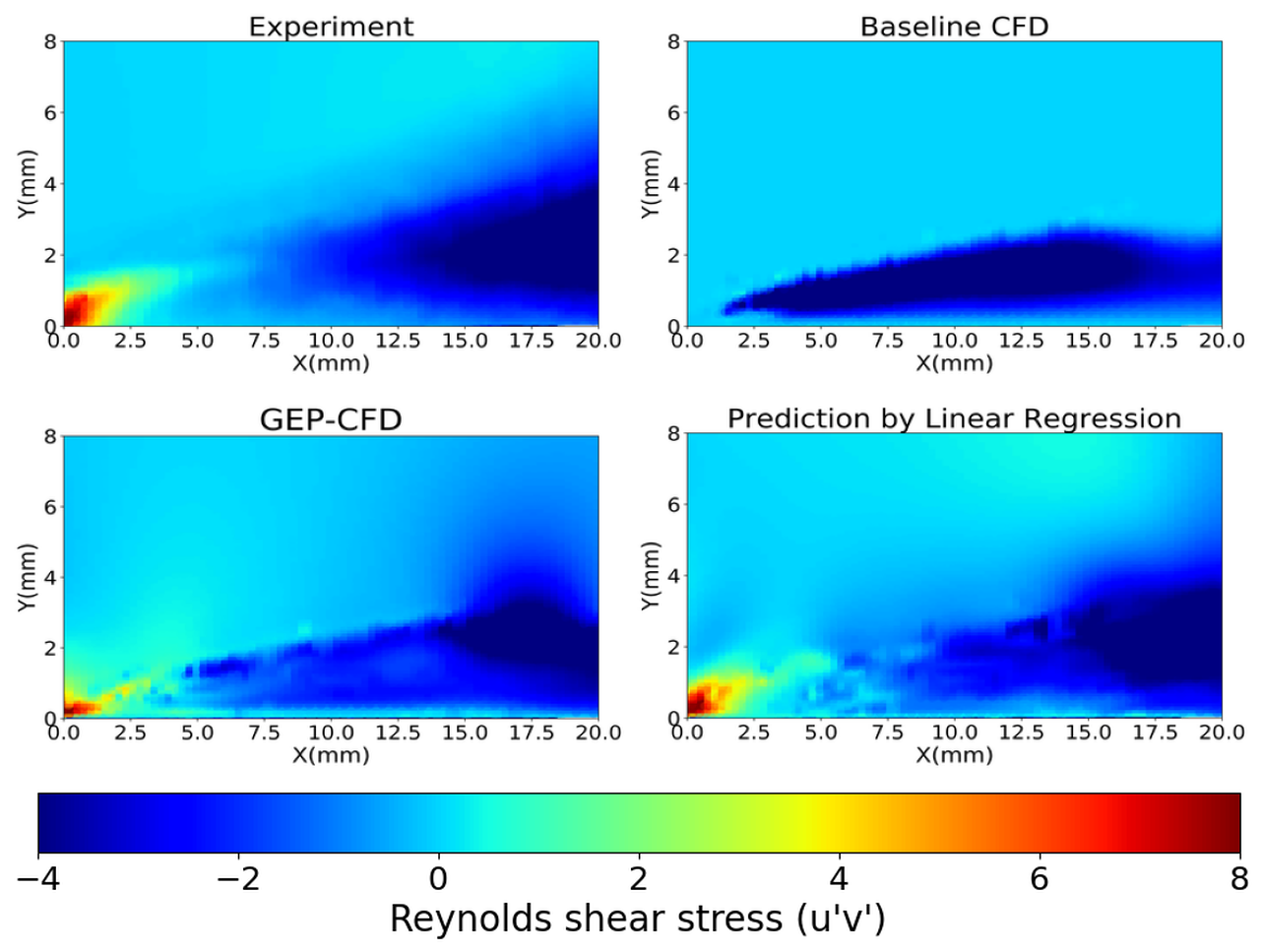}
   \caption{Comparison of results after applying BFGS numerical optimization to the GEP-generated equations. (a) and (b) represent the \textit{u'v'} from experiments and CFD respectively while (c) represents the result after applying numerical optimization. The final figure represents the \textit{u'v'} from linear regression}
  \label{gep_result}
\end{figure}

The mathematical superiority of the optimized expression is shown by the Mean Squared Error (MSE). While the MSE of the best expression derived from the GEP is 1.00851, the MSE of the optimal expression is 0.682. The physics-based superiority of the optimization algorithm is depicted in Fig \ref{gep_result}, which represents the results after applying the numerical optimization as compared to the CFD simulation and experimental data. The optimized result is able to predict the \textit{u'v'} much better than the CFD simulations: it shows high values of negative \textit{u'v'} downstream and high positive values near the throat, in addition to showing the mean cavity shape. A similar observation is noted after applying the linear regression (LR) procedure: its output function, in the form of a seven-degree polynomial is able to reproduce the mean cavity shape and negative \textit{u'v'} downstream, where the rolled-up detached cavity collapses and high positive values near the throat where the cavity is expected to be initiated. However, there is a fundamental difference between the two procedures: while the GEP-optimization approach aims to define a constitutive relation, the LR attempts to simply fit the experimental data, in a much less-complicated fashion. 

\subsection{Scalar Regression for Turbulent kinetic energy}

\begin{figure}[htbp]
    \centering
    \includegraphics[width= \linewidth]{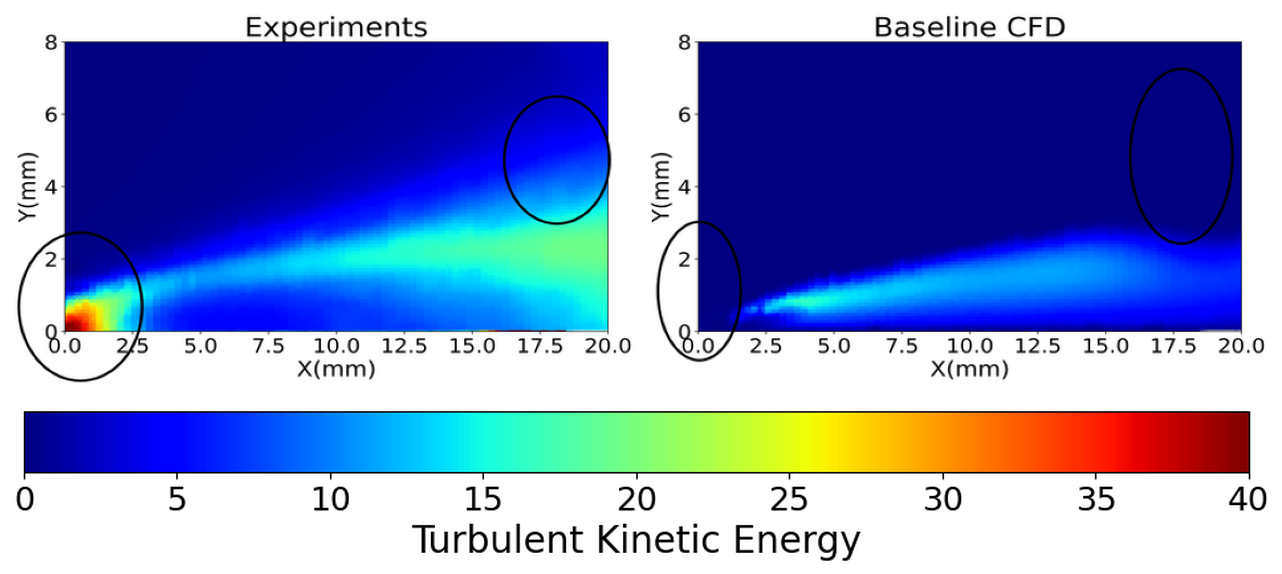}
    \caption{Turbulent kinetic energy for both (a) experiments and (b) CFD simulation with the origin being the venturi's throat. The CFD simulation distinctly under-predicts the TKE throughout the domain producing some TKE where the cloud cavity is expected to form.}
    \label{tke_gep}
\end{figure}

Turbulent kinetic energy is another vital component indicative of flow regime and a paramount variable solved in the two-equation RANS models. The TKE is also associated with the Boussinesq hypothesis in the form of half the sum of the variances of the fluctuating velocity components or the sum of the diagonal components of the Reynolds stress tensor: 
\begin{equation}
TKE = \frac{1}{2}(\bar{u}'^{2} + \bar{v}'^{2} + \bar{w}'^{2})    
\end{equation}
A methodology similar to one described in Section \ref{ss:uv} is performed on the TKE data from CFD simulations to fit the experimental data. Fig \ref{tke_gep} shows the TKE plots for experiments (a) and CFD data (b). In general, the CFD simulation severely under-predicts the TKE, with its inability to predict high TKE in the downstream region, where the cavity detaches into a secondary cavity. In addition, it produces a weakly shaded contour indicating the cloud cavity shape and no high TKE at the venturi throat as well.

Several constitutive relations are derived from EVE for a TKE formulation, as they were generated for \textit{u'v'}.These expressions are then input into an optimized expression  to reduce the uncertainties arising from the parameter changes of the GEP algorithm. The error of the optimized function is reduced using the BFGS algorithm.  Concurrently, the linear regression routine tries to obtain a fit for the experimental data using the variables provided by the URANS simulation. 
\begin{figure}[htbp]
    \centering
    \includegraphics[width=\linewidth]{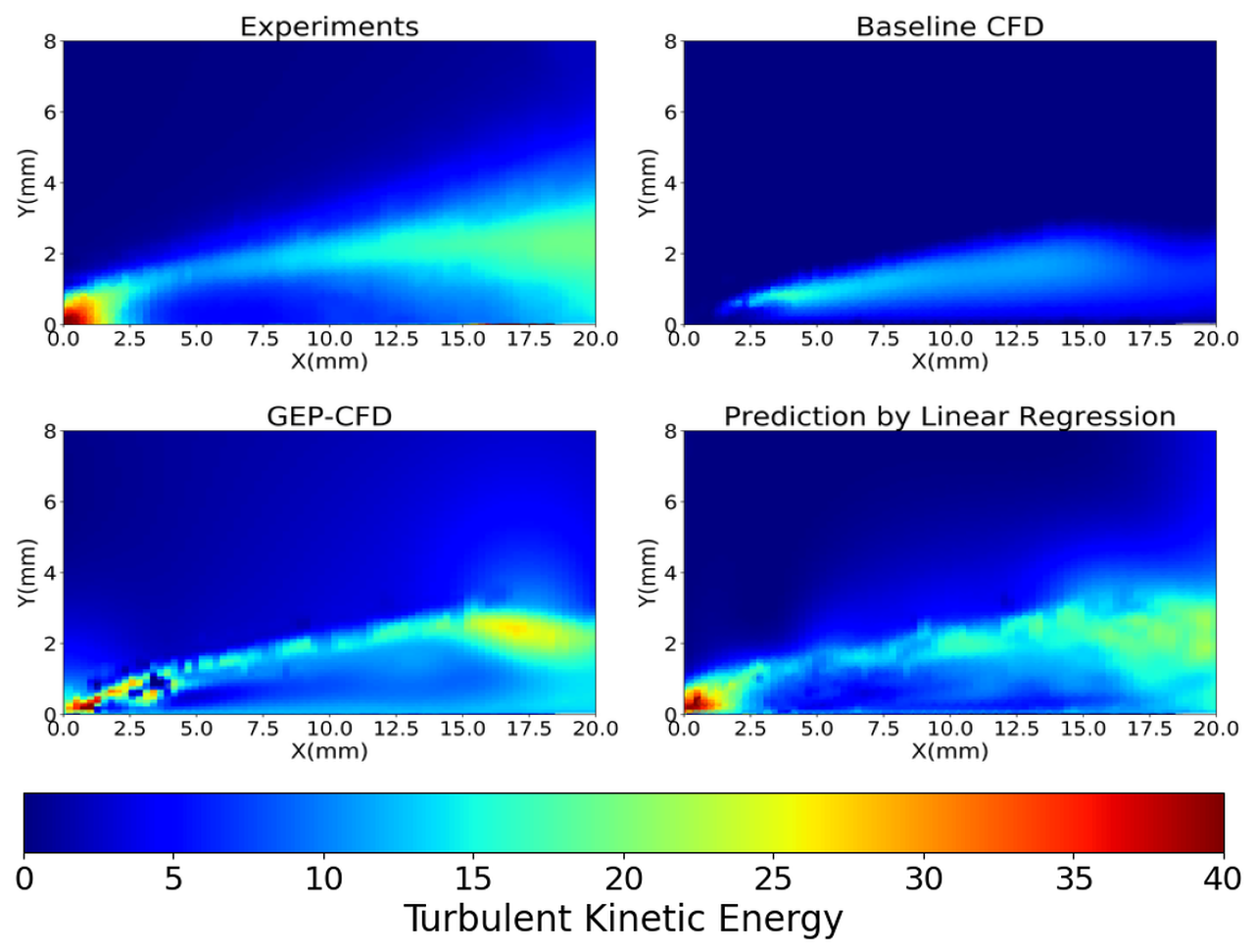}
    \caption{Comparison of results after applying BFGS numerical optimization to the GEP-generated equations. (a) and (b) represent the TKE from experiments and CFD respectively while (c) represents the result after applying numerical optimization. The final figure represents the TKE from linear regression.}
    \label{tke_gep_result}
\end{figure}

Fig \ref{tke_gep_result} shows the TKE plots after the numerical optimization and the linear regression. While the linear regression plot is able to capture the dynamic change observed in the experimental plots, the GEP-optimization procedure is unable to reproduce the high turbulent kinetic energy at the throat and downstream. It is posited that since the TKE data from the URANS simulations utilized in the training data is averaged and does not factor in the velocity fluctuations, the framework is unable to bridge the subsequent differences with the experiments. An additional global field property needs to be incorporated into the framework to ensure the training data includes the fluctuation data. The framework is thus extended to weigh in the standard deviation of velocity in stream-wise and wall direction. The standard deviations indicate veering of the velocity off the mean velocity profiles and therefore, depict the fluctuations in velocity and, the Reynolds stress tensor. Here, the linear regression approach experiences a significant problem - using more variables to obtain the constitutive relation extends the length of the global expression. Previously, the regression-derived expression had 330 terms; it is computationally expensive to input more terms for a single expression. Therefore, the standard deviation data is input only for the GEP data.   
\begin{figure}
    \centering
    \includegraphics[width=\linewidth]{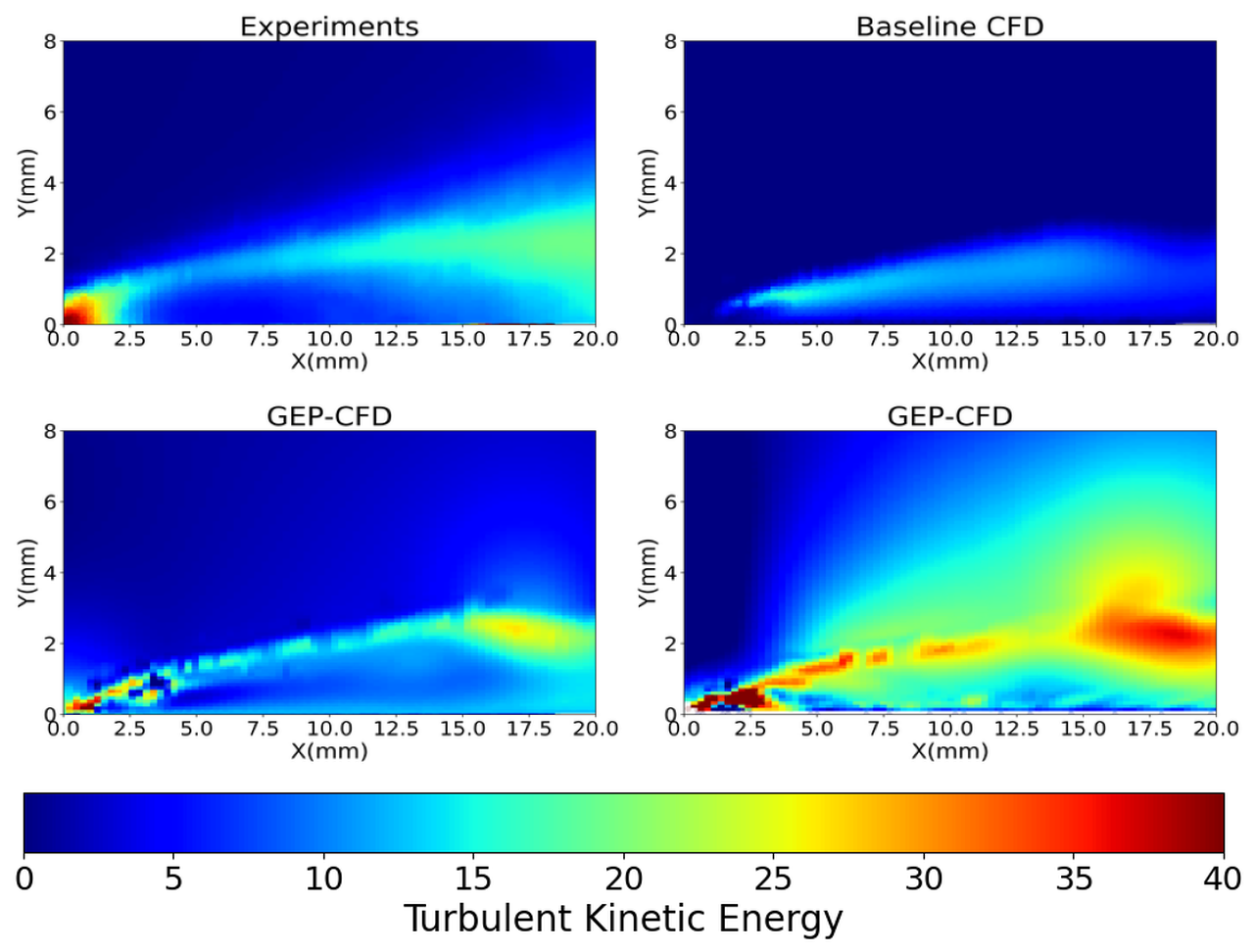}
     \caption{Comparison of results after applying BFGS numerical optimization to the GEP-generated equations. (a) and (b) represent the TKE from experiments and CFD respectively while (c) represents the result before incorporating the standard deviation field values while (d) depicts the TKE after the values are taken into account. The third figure is a duplicate of Fig \ref{tke_gep_result} (c), solely for comparison purposes.} 
\label{tke_gep_result_new}
\end{figure}

The initial successful constitutive relations for TKE developed GEP training runs after incorporating the standard deviation are:
\begin{eqnarray}
\hspace*{-2cm} b_{1,ij} &=& -2\alpha + 2qv + 2q + 1ru - 0.21u + 1x + (1r + 0.15)(1r + 1v) + 4.0  \label{Eq:16}\\
\hspace*{-2cm} b_{2,ij} &=& -0.21\alpha + 2qv + 1q - 1rx + 1r(1r + 1u + 4.0) - 0.43u + 1x + \nonumber \\
\hspace*{-2cm} & & (1r + 0.15)(1v + 2.0) + 5.43 \label{Eq:27} \\
\hspace*{-2cm} b_{3,ij} &=& -2\alpha(0.43u - 2.0) - 1rx + 1r(1qr + 1r + 1u) + 1r + 1v(1q + 2.0) + 1x + \nonumber \\
\hspace*{-2cm} & & (1r - 0.21)(1u - 1v) + 6.1 \label{Eq:28} \\
\hspace*{-2cm} b_{4,ij} &=& 1\alpha x - 2\alpha(0.43u - 1.0) - 1rx + 1r(2.0q + 1r + 1u) + 1v(1q + 2.0) + \nonumber \\
\hspace*{-2cm} & & (1r - 0.21)(1u - 1v) + 8.15\label{Eq:29} \\
\hspace*{-2cm} b_{5,ij} &=& 1r(1q + 1u) - 1v(1\alpha - 1v)(1q + 0.21) + 1x + (1r + 0.43)(1x + 2.0) - \nonumber \\
\hspace*{-2cm} & & (0.31x - 2.1)(2r - 0.15u + 1x + 0.1) + 2.0\label{Eq:30} \\
\hspace*{-2cm} b_{6,ij} &=& 1qv(1\alpha - 1r + 1v) + 1ru(1\alpha + 1r) + 1r(3.0q - 1x) - 0.61u + 0.1v^{2} + 1x + 8.0 \label{Eq:31}
\end{eqnarray}
\begin{figure}[htbp]
    \centering
    \includegraphics[width=\linewidth]{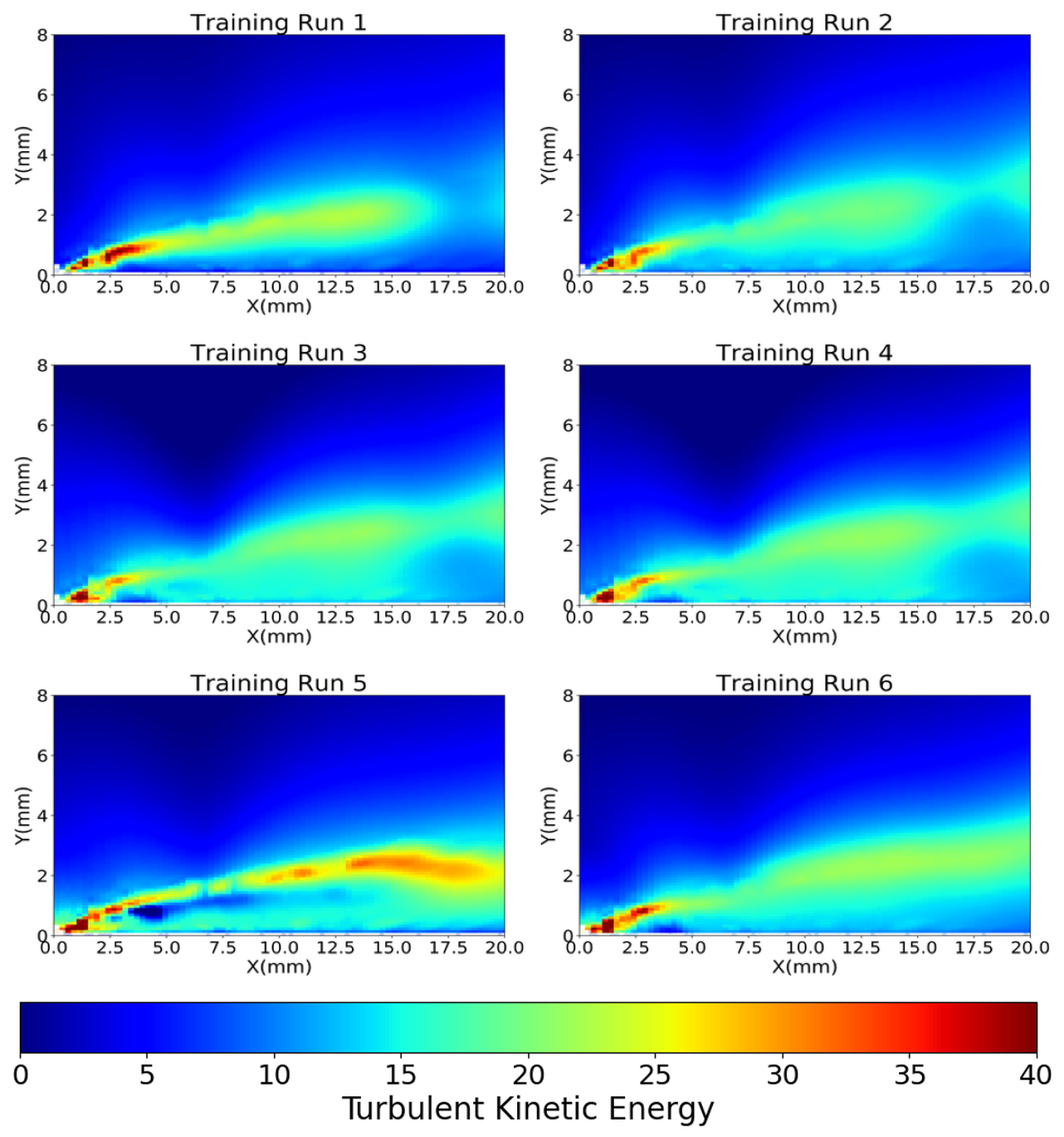}
    \caption{Scatter plots for TKE from five formulations provided from six different GEP training runs with the origin being the venturi throat. All plots except (e) show identical features and are predicting identical turbulence dynamics to CFD data}
    \label{fig:five_gep}
\end{figure}

where \textit{x}, u, v, $\alpha$ are the variables used previously in the previous GEP training runs for TKE. The terms \textit{q} and \textit{r} signify the time-standard deviations of the velocity averaged over time in both stream-wise and wall directions. A set of such expressions are then fed to a BFGS optimizer to produce an optimal expression. Fig \ref{tke_gep_result_new} (d) presents the TKE plot produced by the optimal expression. The prediction becomes much more robust compared to the previous setup. This indicates that the standard deviation of velocities are indeed the deciding factors in the GEP-CFD algorithm for TKE; incorporating the fluctuations from the mean velocity profiles aids the GEP procedure to generate more accurate expressions. Thus, it can be stated that the GEP does understand some flow physics, rather than being a mathematical "blackbox", where the aim is to solely reduce the error function. A discrepancy arises between the experiments and the GEP-CFD approach though: while the GEP-CFD model is able to predict the high TKE at the throat, it also predicts high TKE values at the cavity-water interface and shows much higher TKE values at the cavity's end as compared to experiments.

\subsection{Impact of inputs on Predictive Performance}

This section provides insights into the influence of some input functions and algorithm parameters and their impact on the expressions. Table \ref{tab:uv_inputs} shows the input fields that were employed to derive a constitutive expression for \textit{u'v'}. To evaluate the influence of each field, a sensitivity analysis was performed using a random sampling approach. As stated previously, the expression obtained from the GEP analysis and the subsequent BFGS optimization is a polynomial composed of non-linear and cross-terms between the different input fields. The expression is first decomposed into its constituents and 5000 random samples are drawn for the input fields. These samples are drawn from the pre-determined sample space of the training data.

\begin{table}
    \centering
    \begin{tabular}{cc}
        Variable & Field  \\ \hline
        x & Reynolds Shear Stress calculated from CFD\\
        $\alpha$ & Void Fraction\\
        u & Time-Averaged Velocity in Stream-wise direction\\
        v & Time-Averaged Velocity in Wall direction\\
    \end{tabular}
    \caption{Input fields for GEP to derive \textit{u'v'}}
    \label{tab:uv_inputs}
\end{table}

\begin{figure}[htbp]
    \centering
    \includegraphics[width=\linewidth]{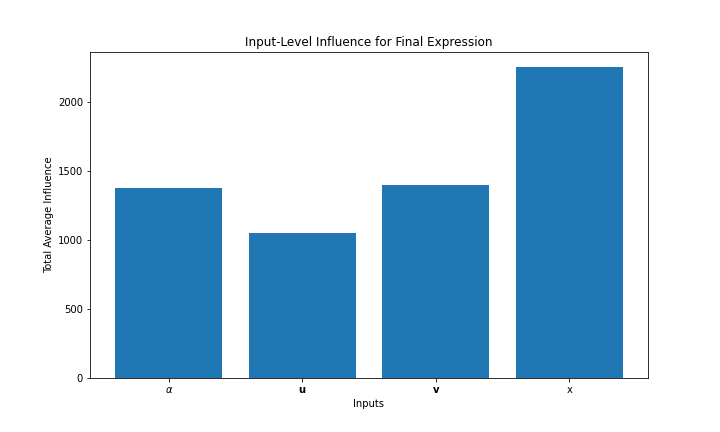}
    \caption{Sensitivity analysis plots for \textit{u'v'}. Here, the Reynolds shear stress from CFD is seen to be the leading influential factor, followed by the void fraction and time-averaged velocity in wall direction, underlining the cavity dynamics and the phase interface respectively. }
    \label{fig:input_influence_uv}
\end{figure}

The influence of each input for \textit{u'v'} was calculated as:
\begin{equation}
    I_{field} = \sum_{t \in \text{Terms where field} \in t} |t(\alpha, u, v, x)|
\end{equation}
where $I_{field}$ is the cumulative influence of each field and \textit{t} represents each component in the polynomial expression. The computed influence was averaged across all the samples to obtain the mean contribution of each field. Fig \ref{fig:input_influence_uv} shows the input-level influence in the optimized expression: the Reynolds stress from CFD simulation (\textit{x}) is predictably, the most impactful field in the expression. Indeed, the framework intends to bridge the gap between the Reynolds shear stress fields between CFD and experiments so it has the highest influence. The two fields, time-averaged velocity in wall direction (v) and void fraction ($\alpha$) are the next important contributing fields. The void fraction shows the cavity dynamics in the simulation and highlights the strong cavitation-turbulence interplay while the time-averaged velocity in wall direction indicates the velocity differences between the vapor and water phases respectively, and thus spotlighting the phase interface.   

To understand how each field individually influences the final expression, independent of its relations with other fields in the final expression, a Partial-Dependence Plot (PDP) analysis is conducted. The PDP of a single field is calculated by varying the field over a pre-defined range and averaging the final expression over all other input fields, sampled from the dataset:
\begin{equation}
    \hat{PDP}(x_{j})= \frac{1}{n} \sum^{n}_{i=1} f_{final}(x_{j}, \textbf{x}^{i}_{-j})
\end{equation}
where $x_{j}$ is the field of interest, $x^{i}_{-j}$ represents all other fields for the \textit{i} th point on the grid and the function $\hat{PDP}(x_{j})$ is the estimated partial dependence, evaluated by averaging over all the samples in the dataset.  Thus, the PDP plots highlight a global view of the expression's behaviour and indicate whether the input has a linear or a non-linear impact on the expression. 

\begin{figure}[htbp]
    \centering
    \includegraphics[width=\linewidth]{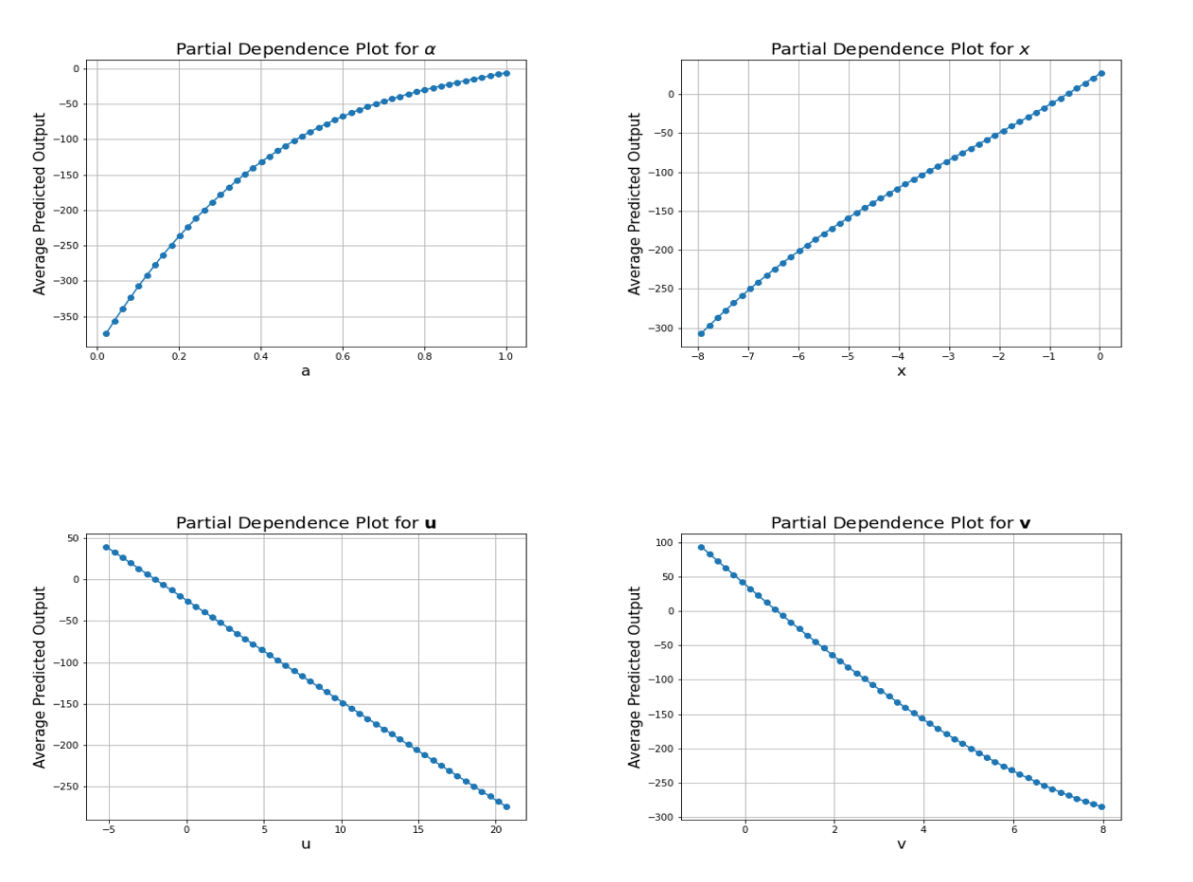}
    \caption{PDP plots for the inputs driving the expression for \textit{u'v'}. The plots indicate the void fraction between 0.2 and 0.8 having significant influence on the final expression along with high Reynolds stress values observed at the cavitating zones and the low time-averaged velocities in both directions. In summary, it can be stated a set of appropriately-defined input fields is able to aid the expression to predict the Reynolds stress accurately.}
    \label{fig:pdp_uv}
\end{figure}

Fig \ref{fig:pdp_uv} represents the PDP plots for each of the input field to drive the final expression. The curve for each field highlights how the field influences the final expression with its own increase in magnitude. For the void fraction ($\alpha$) field, it is observed the value of the expression will increase substantially with increasing values of $\alpha$. An $\alpha$ value ranging between 0.2 and 0.8, indicating a mixture of vapor and water has a significant influence on the expression. For the Reynolds stress from the CFD simulations (\textit{x}), an increase in its magnitude results in a higher magnitude of the final expression. This trend is particularly important near the cloud cavity, where experiments show high Reynolds stress values. These high values impact the final expression's ability to accurately predict Reynolds stress across the entire cavity region. The PDP plots for time-averaged velocities in both directions indicate that an increase in their values adversely affects the magnitude of the final expression. The time-averaged velocities are higher in the regions with pure water, in the upper region of the venturi as compared to the cavitating region. While these quantities influence the expression significantly in the cavitating flow region, they tend to have a minimal influence in the non-cavitating regions. Therefore, the final expression is able to provide an appropriate correction throughout the venturi, including both cavitating and non-cavitating regions. 

\begin{figure}[htbp]
    \centering
    \includegraphics[width=\linewidth]{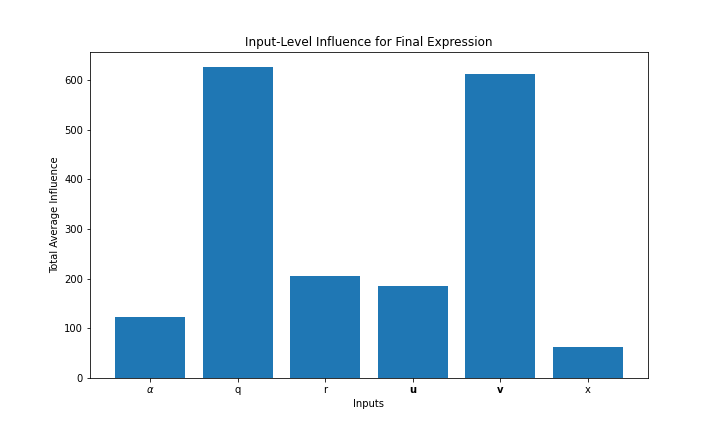}
    \caption{Sensitivity plots for the inputs driving the expression for TKE. The plots indicate the void fraction between 0.2 and 0.8 having significant influence on the final expression along with high Reynolds stress values observed at the cavitating zones and the low time-averaged velocities in both directions. In summary, it can be stated a set of appropriately-defined input fields is able to aid the expression to predict the Reynolds stress accurately.}
    \label{fig:input_influence_k}
\end{figure}

\begin{figure}[htbp]
    \centering
    \includegraphics[width=1.1\linewidth]{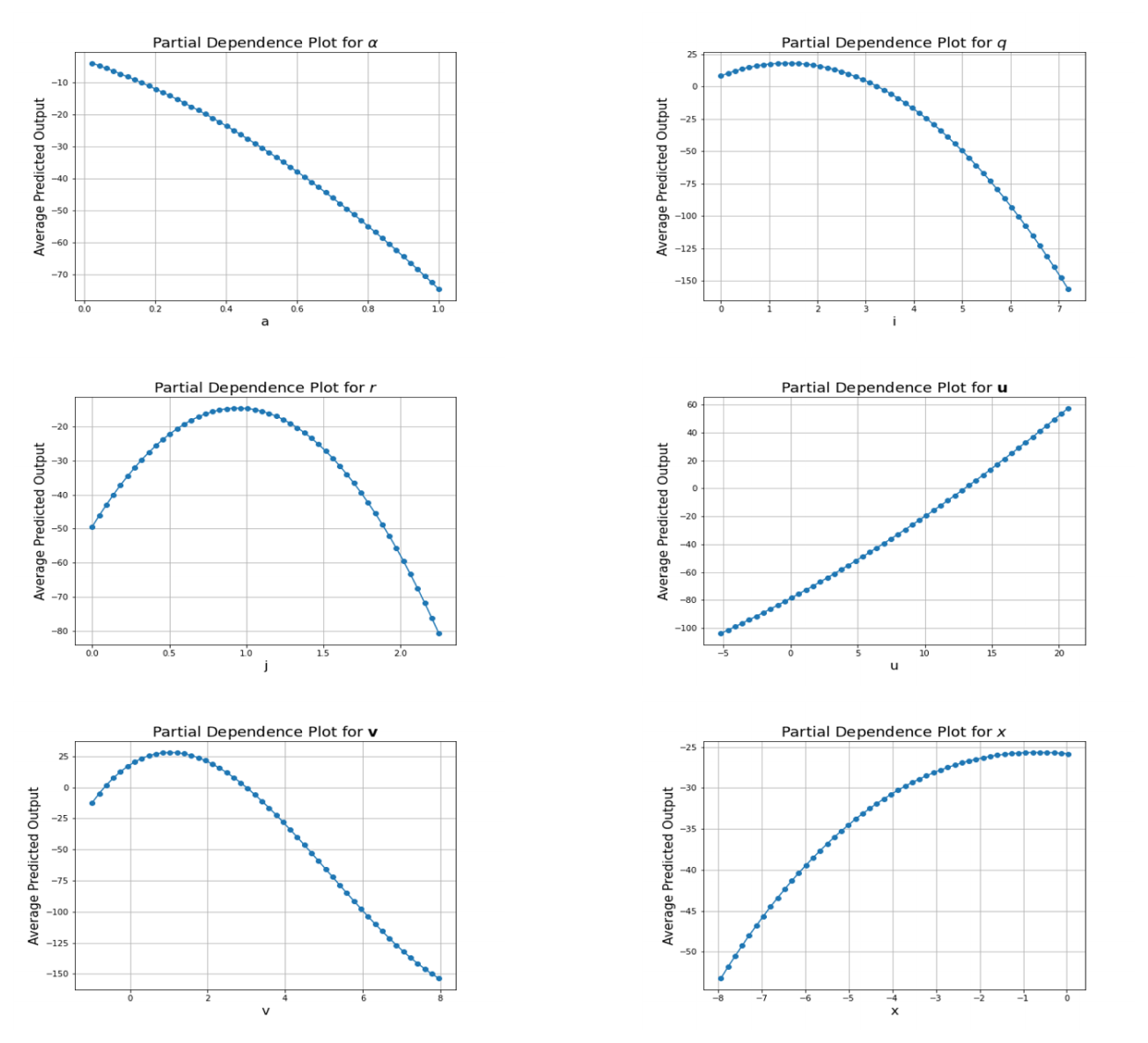}
    \caption{PDP plots for the inputs driving the expression for TKE. The plots indicate a decreasing influence as the void fraction increases but significantly larger influence factors for the standard deviation of velocities in both directions. Additionally, high areas of influence are noted for the range of time-averaged velocities that lie within the cavitating region}
    \label{fig:pdp_k}
\end{figure}

\begin{table}
    \centering
    \begin{tabular}{cc}
        Variable & Field  \\ \hline
        x & Turbulent Kinetic Energy calculated from CFD\\
        $\alpha$ & Void Fraction\\
        u & Time-Averaged Velocity in Stream-wise direction\\
        v & Time-Averaged Velocity in Wall direction\\
        q & Time-Averaged standard deviation of Velocity in Stream-wise direction\\
        r & Time-Averaged standard deviation of Velocity in Wall direction\\
    \end{tabular}
    \caption{Input fields for GEP to derive TKE}
    \label{tab:k_inputs}
\end{table}
Fig \ref{fig:input_influence_k} depicts the input-level influence for the final expression for TKE: the top influential input fields are the time-averaged velocities in both directions (\textbf{u,v}) and their standard deviations (\textit{q,r}), especially the standard deviation of the time-averaged velocity in streamwise direction. This reaffirms the influence of the standard deviation fields on the final expression and the fact that GEP does understand some flow physics. On the other hand, the void fraction ($\alpha$) and TKE from the CFD simulation have a comparatively lesser influence on the final expression. 
Fig \ref{fig:pdp_k} represents the PDP plots and the individual relationships between the final expression and the input fields. The void fraction field ($\alpha$) plot shows that an increase in void fraction results in a decrease in the value of the final expression. The observation can be noted with regards to fig \ref{tke_gep_result_new} where higher values of $\alpha$ indicate non-cavitating regions where the TKE field values are low. The standard deviation for both velocity fields indicate an adverse relationship with the expression magnitude as well, although a slight anomaly is observed with the standard deviation of velocity in wall direction (\textit{r}) where the final expression magnitude initially increases with increase in the input field value. The plots show the presence of fluctuations influencing the final expression significantly until they reach physically unfeasible values. The time-averaged velocity in stream-wise direction (u) has linearly increasing relationship with the final expression while the time-averaged velocity in wall direction (v) initially has slightly upward peak before advancing to an adverse impact on the final expression's magnitude. However, the range of these input fields for the cavitating zone lies within the range where both fields have a positive relationship with the final expression. Once the input fields exceed the range, effectively in the non-cavitating region, the value of the final expression decreases markedly to zero. The PDP plot for the TKE from CFD simulations (\textit{x}) shows a non-linear relationship where an increase in \textit{x} results in an increase in the value of the final expression. The relationship enhances prediction because, even though the CFD may under-predict the TKE field, it can still identify the cavitating zone by showing relatively higher TKE values. This helps the final expression to predict high TKE  more accurately in the zone. In summary, the strong influence of the void fraction to drive the final expression for \textit{u'v'} underscores the strong cavitation-turbulence interplay while the time-averaged velocity field in wall direction highlights the importance of the phase interface for the final expression of \textit{u'v'}. Similarly, the significant influence of the standard deviation fields of the time-averaged velocities on the final expression for TKE is noted along with the influence of other inputs fields within the range of values belonging to the cavitating zone. These factors collectively enhance the accuracy of the final expression to predict the TKE field values. 

\subsection{Parameter Importance analysis for algorithm parameters}

\begin{figure}[htbp]
    \centering
    \includegraphics[width=1.1\linewidth]{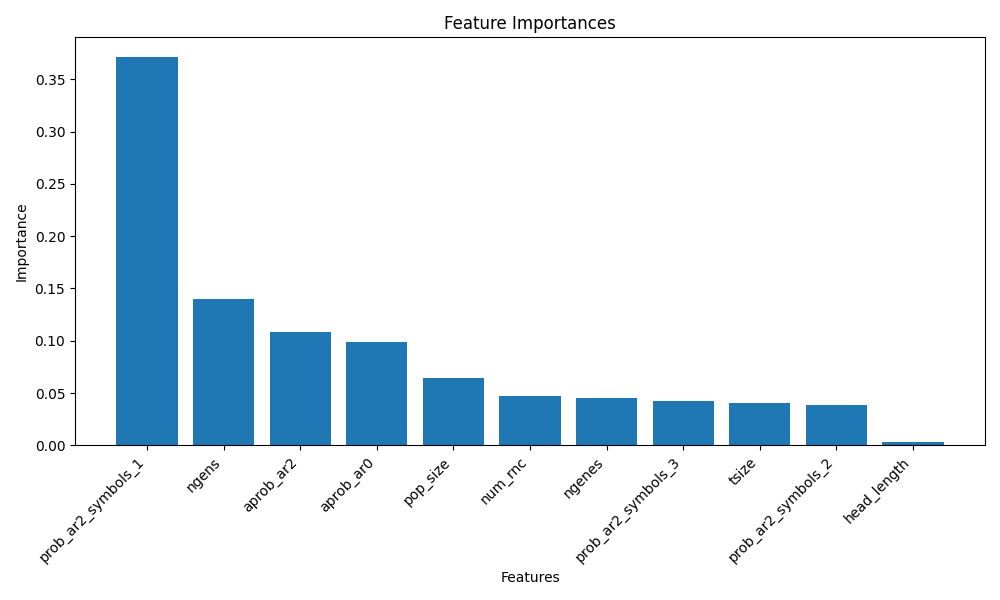}
    \caption{Feature importance graphs for the parameters varied for difference expression. The probability of symbol with arity 2 is the highest, especially the multiplicative identity [$\times$] (prob\textunderscore ar2\textunderscore symbols\textunderscore 1) followed by the number of generations. The other parameters have considerably lesser importance factors}
    \label{fig:feature_importance_uv}
\end{figure}

Both the input fields and the algorithm parameters, or features influence the expressions generated by GEP and the subsequent optimized expression. It is therefore imperative to evaluate the importance of the features as well. An ensemble learning method, titled as Random Forest is applied for the parameters driving the expressions with the highest influence on the final expression. This influence is noted using the absolute values of their optimized coefficients. Multiple decision trees are constructed with the parameter's importance quantified by the extent it reduces the prediction error during a tree-splitting node. The importance is calculated as:
\begin{equation}
    Importance (param) = \frac{1}{T} \sum_{t=1}^{T} \sum_{n \in nodes(t)} I(param) \Delta Error (n)
\end{equation}
where T is the total number of trees, nodes(t) is the set of all nodes in tree \textit{t}, I(param) is an indicator function that equals 1 if the node n uses the parameter, \textit{param} and 0 otherwise while $\Delta Error (n)$ represents the reduction in error achieved by splitting the data at node n using the parameter \textit{param}. 
The reduction in error is computed as:
\begin{equation}
    \Delta Error (n) = \text{Error before split} - (\frac{N_{left}}{N} \times Error_{left} + \frac{N_{right}}{N} \times Error_{right})
\end{equation}
where N is the total number of samples at node \textit{n}, $N_{left}$ and $N_{right}$ are the number of samples in left and right child nodes, respectively. The \textit{Error} indicates the Mean Squared Error (MSE). The error reductions contributed by each parameter across all trees in the Forest are summed and the resulting values are normalized. The parameters with higher normalized scores contribute frequently and significantly to the development of the expression. Fig \ref{fig:feature_importance_uv} shows the importance factors for various parameters that drive the expressions for \textit{u'v'}. The most influential parameter is the probability of the symbols having arity of two, specifically the multiplication symbol (prob\textunderscore ar2\textunderscore symbols\textunderscore 1). The other two symbols with the same arity, addition and subtraction (prob\textunderscore ar2\textunderscore symbols\textunderscore 2 and prob\textunderscore ar2\textunderscore symbols\textunderscore 3 respectively) have considerably lower importance factors. As the probability of a multiplicative index increases, the likelihood of more expressions satisfying its arity also increases. The \textit{ngens} parameter, the number of generations before the GEP stops is also an influential parameter as more generations allow for more candidate solutions to be generated. There are several other factors as listed however, the above-described ones yield the most influence on the GEP expressions, and subsequently the optimized expression. 
\subsection{Multi-expression Optimization for Turbulence Closure }

The previous methodologies specify training and optimizing a single constitutive relation, either \textit{u'v'} or TKE. However, as stated previously, optimizing a sole expression is theoretically infeasible and the objective is to train both expressions simultaneously. Waschkowski et al. \cite{waschkowski2022multi} devised multi-expression framework where multiple closure models were generated simultaneously in a training run. The individual closure model relations are now grouped into colonies. Since multiple objectives are of interest, the fitness value will be calculated as a weighted sum of the fitness values of the individual expression and genetic material will be exchanged only between individuals with same closure model relation. Each colony consists of two individual expressions, one for \textit{u'v'} and one for TKE. Thereafter, a population of such colonies is used in multi-expression optimization rather than solely an individual closure model relation. It is important to emphasize here that the genetic material and expression properties are exchanged only between individuals of the same closure model. Here, the net fitness function is calculated as sum of the fitness functions for both turbulent quantities without any specific weight values. The weights have not been assigned to ensure both objectives are equally satisfied by each colony rather than satisfying a singular objective turbulence field.  The GEP procedure is conducted for both quantities to draw several expressions, which are fed into the BFGS optimizer function to design an optimal constitutive relation. Here, the linear regression routine is unable to predict two expressions with multiple objectives simultaneously and is therefore not employed.

The first few successful training runs for the multi-expression training are:
The initial successful constitutive relations for TKE developed GEP training runs after incorporating the standard deviation are:
\begin{eqnarray}
\hspace*{-2cm} f_{1,ij} &=& -0.09r(1\alpha - 2q + 1r - 0.43u(1q + 1v) + 3u - 1x + (1\alpha - 1v)(1u + 1v)) \\
\hspace*{-2cm} b_{1,ij} &=& 1q(1r + 1v(1v - 0.15)) + 1qu + 3r - 0.21u(1\alpha - 1r + 0.43) + 1m + 3.58  \\
\hspace*{-2cm} f_{2,ij} &=& -(1r - 0.15x + 0.07)(0.22u - 1.45v + 0.25) \\
\hspace*{-2cm} b_{2,ij} &=& 1\alpha^{2}m(1\alpha - 1q + 0.18) + 1qv(1v - 0.15)(1\alpha - 1r + 1.0) + 4.0q + 1r + 1m + \nonumber \\
\hspace*{-2cm} & & (1r - 0.21)(1u + 2.0) + 1.43 \\
\hspace*{-2cm} f_{3,ij} &=& -2\alpha r + 1qr - 1r - (1r + 0.15)(1\alpha + 1r - 1v) \\
\hspace*{-2cm} b_{3,ij} &=& 1q + 1r^{2}(1q + 1r + 2u -1m + 2.0) + 2.0rv^{2} - 0.21u + 1m + \nonumber \\
\hspace*{-2cm} & & (1\alpha - 0.09)(1q + 0.21) + 2.0  \\
\hspace*{-2cm} f_{4,ij} &=& -1\alpha - 1r(2\alpha - 0.09)(0.09\alpha(1\alpha - 1r) + 0.91) + 1v(1\alpha q + 0.21) \\
\hspace*{-2cm} b_{4,ij} &=& 1q(1r + 1v - 0.15m + 2.0) + 1q + 1u(-2\alpha + 1r + 1.9) + 0.09v + 1.09m \\
\hspace*{-2cm} f_{5,ij} &=& -1\alpha r(1\alpha - 1v) - (1\alpha - 1r)(0.43\alpha + 0.43q - 0.43x - 0.1) \\
\hspace*{-2cm} b_{5,ij} &=& 1\alpha y + 1q + 1ru + 1r(1q + 2r + 2.0) + 1v(0.43v + (1\alpha - 1r)(1m - 2.0)) + 1.1 \\
\hspace*{-2cm} f_{6,ij} &=& (0.43x - 0.21)(1\alpha - 0.04q + 0.21u - 1.21v + 0.21x - 0.03) \\
\hspace*{-2cm} b_{6,ij} &=& 1r(1u(1r + 0.43) + 2.0) + 1v(2.0q - 0.15u + 1v - 1.0) + \nonumber \\
\hspace*{-2cm} & & (1r + 4.0)(1r - 0.15u + 3.0) \nonumber \\
\end{eqnarray}
where $f_{n,ij}$ and $b_{n,ij}$ represent the closure expressions for \textit{u'v'} and TKE respectively, \textit{m} denotes the TKE field values from the baseline CFD simulations and \textit{x, q, r, u, v}, $\alpha$ representing the variables as used for the previous GEP training runs. 
\begin{figure}[htbp]
    \centering
    \includegraphics[width=\linewidth]{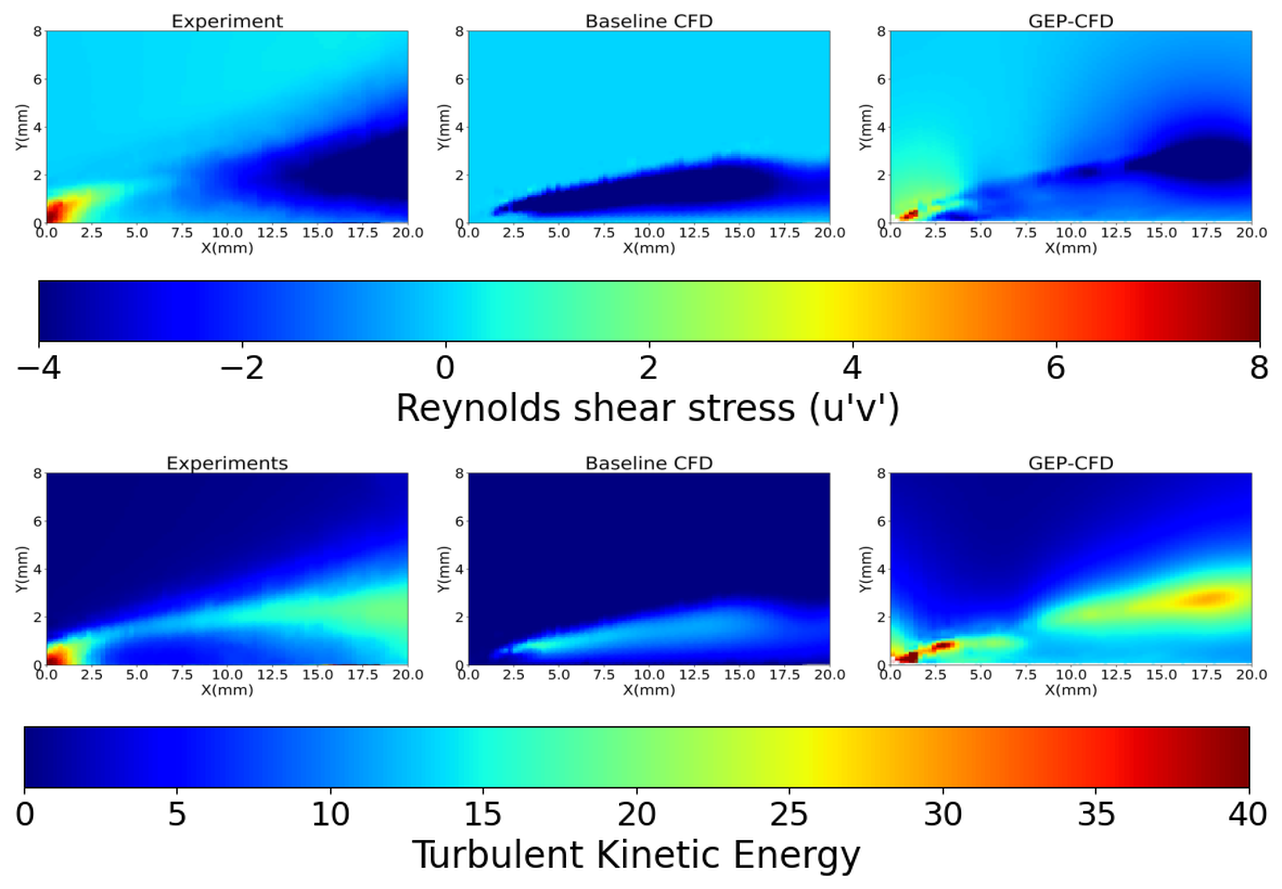}
    \caption{Plots showing both \textit{u'v'} and TKE from multi-expression optimization. Subfigures (a), (d) represent the experimental plots, (b), (e) show the CFD plots followed by (c) and (f) from optimization. The framework outperforms the CFD calculations on both aspects.}
    \label{fig:multi}
\end{figure}

\begin{figure}[htbp]
    \centering
    \includegraphics[width=\linewidth]{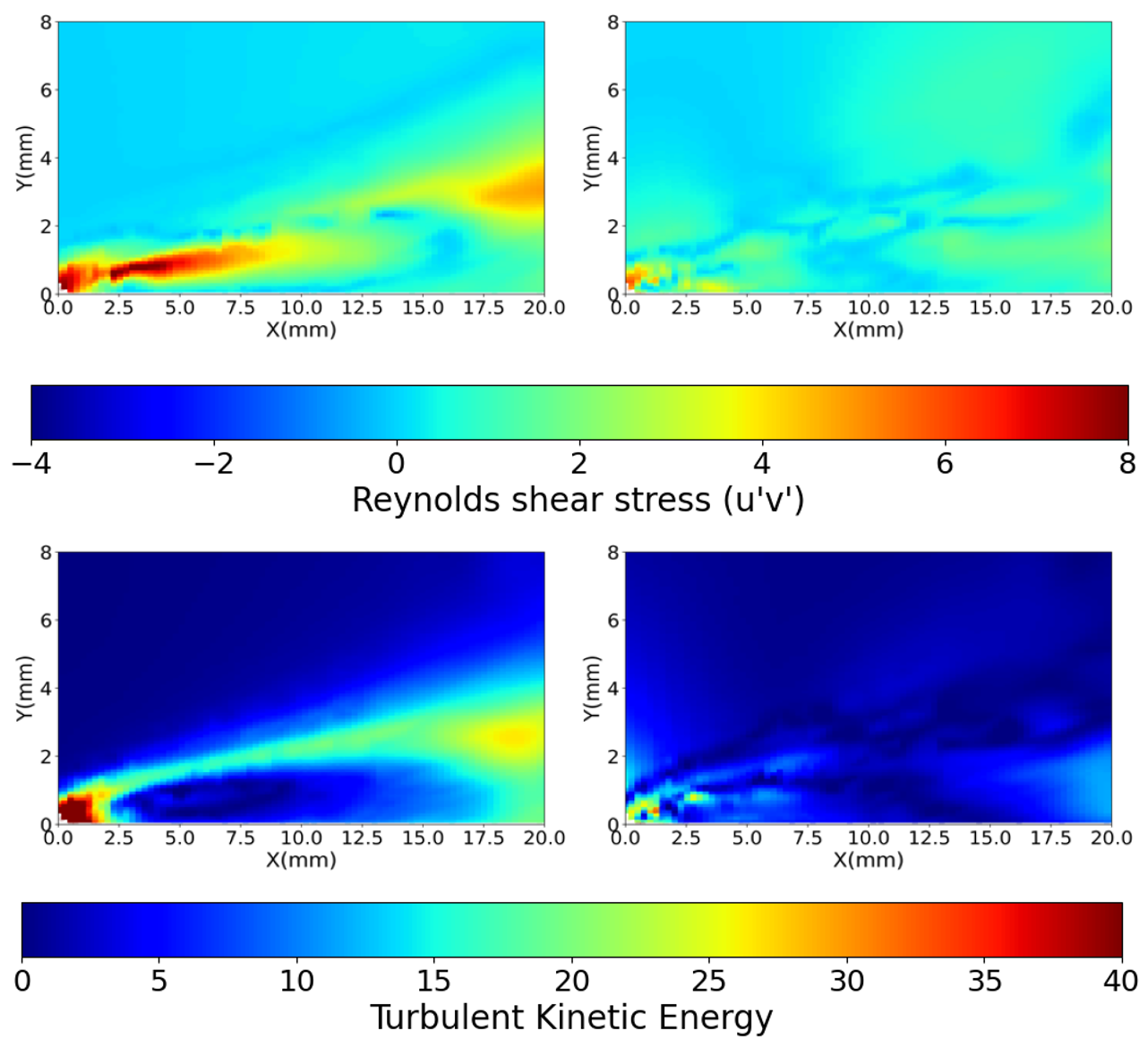}
    \caption{Error Plots signifying the accuracy of the GEP procedure. Subfigures (a) and (b) show the errors between CFD results \& experimental data and GEP optimization \& experimental data for \textit{u'v'} while (c) and (d) show the errors between CFD results \& experimental data and GEP optimization \& experimental data for TKE}
    \label{fig:multi_error}
\end{figure}
Fig \ref{fig:multi} shows the plots drawn by the optimized constitutive expressions for both turbulence properties. The first row denotes the plots for \textit{u'v'}. The optimization is able to predict the high value of \textit{u'v'} near the throat where cavitation inception occurs, albeit with less magnitude as compared to the experiment. Similarly, in the downstream region away from the wall, the high magnitude \textit{u'v'} region is under-predicted only slightly as compared to the CFD model where, the turbulence at the cavity detachment is not captured. Extending the comparison to TKE also yields identical results: the multi-expression optimization is able to well-predict the magnitude of TKE field throughout the \textit{venturi} nozzle and is able to capture the flow features close to the throat and further downstream. Moreover, the multi-expression GEP-CFD approach predicts high TKE values only near the throat and the edge of the cavity, similar to the experimental data. This observation differs sharply from the single expression GEP-CFD approach for TKE, where it predicted high TKE throughout the cavity as well. The results are affirmed in Fig \ref{fig:multi_error} which depicts the errors between the CFD simulations \& the experiments and the GEP optimization \& the experimental data for both fields. The subfigures (a) and (c) highlight the large discrepancies shown by CFD simulations for \textit{u'v'} and TKE, especially in the cavity formation where it under-predicts the cavity. On the other hand, the multi-expression GEP-CFD approach has comparatively fewer errors throughout the vapor cavity region.

Indeed in both turbulence dynamics expressions, the GEP-developed constitutive relations outperforms the baseline CFD relations with their ability to predict the high cavitation-turbulence interactions near the throat and downstream where the detached cavity rolls up and collapses. The errors exhibited by the GEP-developed constitutive relations are consistently lower than the baseline CFD simulations. 

\section{Conclusions}\label{se:pap3_conclusions}

In this study, the concept of Gene-Expression Programming (GEP) was applied to formulate closure expressions for the Reynolds shear stress and Turbulent Kinetic Energy (TKE) and account for the discrepancies posed by the Boussinesq approximation for URANS modelling in cavitating flows. To the best knowledge of the authors, this is the first application of GEP for developing closure expressions for turbulent, unsteady cavitating flows. The closure expression, initially generated for Reynolds stress used a GEP framework, where randomized individual expressions, developed to fit the dataset, are created and made to compete in tournaments so that only the fittest expressions qualify for the next round. The expressions were defined as a function of flow features that include time-averaged velocities in stream-wise and wall directions, void fraction and the Reynolds stress calculated from the Boussinesq approximation  obtained from baseline URANS simulations to ensure the algorithm is able to account for some flow physics. 

Furthermore, the uncertainties manifested by the algorithm parameters were then reduced by feeding several GEP-generated expressions into a BFGS optimizer to compose a single optimal expression. The optimal expression outperformed traditional URANS models and its individual component GEP functions, and was also compared against a simple linear regression routine to evaluate its need for complexity. A similar methodology was devised for the TKE and led to some similar results; however, additional parameters like standard deviation of velocity had to be taken into consideration to visualize the flow features accurately. This was confirmed additionally, by further expression analysis conducted to comprehend the importance of various input fields and the GEP algorithm parameters. Here, Partial Dependence Plots (PDP) show the standard deviation input fields have considerable influence on the expression for TKE followed by the time-averaged velocities. On the other hand, the expression for \textit{u'v'} is driven to a major extent by the void fraction field and the time-averaged velocity in the wall direction, highlighting the influence of cavity dynamics and underlining the cavitation-turbulence interaction. 

Secondly, a joint approach was conducted where the candidates for both the turbulence properties were coupled to obtain a set of joint optimized expressions. Comparisons against CFD simulations showed the multi-expression GEP-optimization framework was able to predict the cavitation-turbulence interplay much better than URANS models and single-expression GEP-optimization framework and reacted positively to including more physical parameters than linear regression.

This work sets further avenues of future research as well. Future studies will focus on conducting a live correction of the Reynolds stress tensor in a cavitating flow. While the idea to plug in the GEP framework into a CFD solver and run successive CFD simulations based on their increasingly fit expression to obtain a final accurate CFD simulation seems possible in fundamental cases, the intense computing time and complexity taken to run a cavitating flow calculation makes this process cumbersome and computationally expensive. Thus, fitting the corrective term directly into the Reynolds stress term periodically during a live simulation is a reasonable prospect. Future studies could drive towards implementing the expression directly into the solver's momentum equation and test the efficiency of this corrective term for other cavitating flow cases as well. In addition, further investigation will be conducted to define the underlying flow physics behind the corrective terms in the new constitutive relations for the turbulence fields. A second avenue of research will be reconstructing the closure relations for the entire Reynolds stress tensor in multi-expression optimization. In this case, weighting of multiple objective functions will be challenging and capabilities for candidate solutions to hold multiple fitness values could aid the process.

\section{Funding}
This work was supported by the Office of Naval Research, USA [grant number N00014-18-S-B001]. The authors would like to thank the ONR proposal manager Dr. Ki-Han Kim for his support.

\section{Data Availability}

The data that support the findings of this study are available from the corresponding author upon reasonable request.

%\section{Bibliography}
\bibliographystyle{elsarticle-num} 
\bibliography{references}
\end{document}